\documentstyle[12pt,aip]{article}

\author{A.P. BALACHANDRAN\\
Physics Department, Syracuse University\\
Syracuse, NY  13244-1130}
\title{GAUGE SYMMETRIES, TOPOLOGY AND QUANTISATION\footnote{Lectures delivered
in the Summer Course on ``Low Dimensional Quantum Field Theories for Condensed
Matter Physicists'', International Centre for Theoretical Physics, Trieste, 24
August to 4 September, 1992.}
}
\date{SU-4240-506\\
September 1992
}
\input tcilatex

\QQQ{Language}{
American English
}

\begin{document}

\maketitle
\begin{abstract}
The following two loosely connected sets of topics are reviewed in these
lecture notes: 1) Gauge invariance, its treatment in field theories and its
implications for internal symmetries and edge states such as those in the
quantum Hall effect. 2) Quantisation on multiply connected spaces and a
topological proof the spin-statistics theorem which avoids quantum field
theory and relativity. Under 1), after explaining the meaning of gauge
invariance and the theory of constraints, we discuss boundary conditions on
gauge transformations and the definition of internal symmetries in gauge
field theories. We then show how the edge states in the quantum Hall effect
can be derived from the Chern-Simons action using the preceding ideas. Under
2), after explaining the significance of fibre bundles for quantum physics,
we review quantisation on multiply connected spaces in detail, explaining
also mathematical ideas such as those of the universal covering space and
the fundamental group. These ideas are then used to prove the aforementioned
topological spin-statistics theorem.
\end{abstract}
\tableofcontents

\newpage

\section{INTRODUCTION}

In recent years, there have been several important developments in low
dimensional quantum physics such as those associated with conformal and
Chern-Simons field theories, the quantum Hall effect and anyon physics.
These lecture notes will address certain aspects of these developments, in
particular those concerning gauge invariance and multiple connectivity and
their consequences for low dimensional physics.

The material in these notes is organised as follows. In Chapter 2, we
discuss the meaning of gauge symmetries and their distinction from
conventional symmetries in general terms. The reason why gauge invariance
leads to constrained Hamiltonian dynamics is also pointed out using
qualitative arguments. An important tool for the quantisation of theories
with constraints is the Dirac-Bergmann theory of constraints and that is
briefly reviewed in Chapter 3.

Chapters 4 and 5 deal with important technical aspects regarding the
treatment of constraints in gauge field theories and some of their physical
consequences. The intimate and beautiful relationship between symmetries and
gauge invariance is clarified and the general theory illustrated by examples
from electrodynamics and the quantum Hall effect. The relation of the edge
states and source excitations in the Hall system to gauge invariance is in
particular explained in Chapter 5.

The remaining Chapters deal with quantisation of classical theories in
multiply connected configuration spaces. As indicated previously, this topic
has assumed importance in low dimensional physics. It has an especially
crucial role in Hall effect and anyon physics where fractional statistics
has a basic significance, statistics being a manifestation of configuration
space connectivity. The notes conclude with a proof of the spin-statistics
theorem in Chapter 7 using topological methods. This proof avoids the use of
relativistic quantum fields and seems well adapted to condensed matter
systems where such fields are not generally of any relevance.

\section{MEANING\ OF\ GAUGE\ INVARIANCE}

The subject matter of our first few Chapters is gauge invariance and its
physical implications. \ We will introduce the topic of gauge
transformations in this Chapter, discussing it in general conceptual terms
[following ref. 1] and emphasizing its distinction from ordinary (global)
symmetry transformations.

\subsection{The Action}

The action $S$ is a functional of fields with values in a suitable range
space. The domain of the fields is a suitable parameter space.

Thus for a nonrelativistic particle, the range space may be ${\bf R}^3$, a
point of which denotes the position of the particle. The parameter space is $%
{\bf R}^1$, a point of which denotes an instant $t$ of time. The fields $q$
are functions from ${\bf R}^1$ to ${\bf R}^3$. Thus, if $F({\bf R}^1,{\bf R}%
^3)$ is the collection of these fields,
\begin{equation}
F({\bf R}^1,{\bf R}^3)=\{q\},\quad q=(q_1,q_2,q_3),q(t)\in {\bf R}^3.
\end{equation}
In other words, each field $q$ assigns a point $q(t)$ in ${\bf R}^3$ to each
instant of time $t$.

For a real scalar field theory in Minkowski space $M^4$, the parameter space
is $M^4$, the range space is ${\bf R}^1$ and the set of fields $F({\bf R}^4,%
{\bf R}^1)$ is the set of functions from ${\bf R}^4$ to ${\bf R}^1$.

Let us denote the parameter space by $D$, the range space by $R$ and the set
of fields by $F(D,R)$. Then the action $S$ is a function on $F(D,R)$ with
values in ${\bf R}^1$. It assigns a real number $S(f)$ to each $f\in F(D,R)$%
. For instance, in the nonrelativistic example cited above,

\begin{equation}
S(q)=\frac m2\int dt\frac{dq_i(t)}{dt}\frac{dq_i(t)}{dt}.
\end{equation}
[The action also depends on the limits of time integration. Since these
limits are not important for us, they have been ignored here. If necessary,
they can be introduced by restricting $D$ suitably. In this case, for
example, instead of ${\bf R}^1$, we can choose the interval $t_1\leq t\leq
t_2$ for $D$.]

The concept of a {\it global symmetry group} may be defined as follows:
Suppose $G=\{g\}$ is a group with a specified action $r\rightarrow gr$ on $%
R\equiv \{r\}$. Then, $G$ has a natural action $f\rightarrow gf$ on $F(D,R)$%
, where $(gf)(t)=gf(t)$. This group of transformations on $F(D,R)$ is the
global group associated with $G$. We denote it by the same symbol $G$. We
say further that $G$ is a {\it global symmetry group} if

\begin{equation}
S(f)=S(gf)
\end{equation}
up to surface terms. For simplicity, we will assume hereafter that $G$ is a
connected Lie group.

As an example, consider the nonrelativistic free particle with $D=\{t\mid
-\infty <t<\infty \}$, $R={\bf R}^3$ and $G=SO(3)$. The rotation group has a
standard action on ${\bf R}^3$. It can be ``lifted'' to the action $%
q\rightarrow gq$ on $F({\bf R}^1,{\bf R}^3)$, where

\begin{equation}
[gq](t)=gq(t)\quad [\equiv (g_{ij}q_j(t))]\ .
\end{equation}
Thus in the usual language, $g$ is a global rotation. Further, $SO(3)$ is a
global symmetry group since for (2.4),

\begin{equation}
S(q)=S(gq)\ .
\end{equation}

In contrast, the {\it gauge group }$\TeXButton{G}{\hat{\cal G}}$ {\it %
associated with} {\it a global group} $G${\it \ }is defined to be the set of
all functions $F(D,G)=\{h\}$ from $D$ to $G$ [with a group composition law
to be defined below]. An element $h$ of $F(D,G)$ thus assigns an element $%
h(d)$ of $G$ for each point $d$ in $D$:

\begin{equation}
D\ni d\stackrel{h}{\rightarrow }h(d)\in G.
\end{equation}
[The hat for $\TeXButton{G}{\hat{\cal G}}$ is put there to distinguish it
from ${\cal G}$ which will occur later.] The group multiplication in $
\TeXButton{G}{\hat{\cal G}}$ is defined by $(hh^{\prime })(d)=h(d)h^{\prime
}(d)$. This group as well has a natural action $f\rightarrow hf$ on $F(D,R)$
defined by $(hf)(d)=h(d)f(d)$. If $S$ is invariant under $\TeXButton{G}
{\hat{\cal G}}$ (up to surface terms), that is, if $S(hf)=S(f\bar )+$%
possible surface terms, then the gauge group is a {\it gauge symmetry group}.

It is possible that the sort of boundary conditions we impose on the set of
functions in the gauge group can have serious consequences for the theory as
we shall see in Chapter 4. See also ref. 2.

Let $\TeXButton{G}{\hat{\cal G}}$ be a gauge symmetry group and let $\Gamma $
be a global symmetry group where $\TeXButton{G}{\hat{\cal G}}$ is not
necessarily associated with $\Gamma $. Recall that the parameter space
contains a coordinate which we identify as time $t$. {\it The profound
difference between} $\TeXButton{G}{\hat{\cal G}}$ {\it and} $\Gamma $ {\it %
is due to the fact that}${\cal \ }\TeXButton{G}{\hat{\cal G}}$ {\it contains
time dependent} {\it transformations unlike} $\Gamma $. It affects the
deterministic aspects of the theory and also has its impact on Noether's
derivation of conservation laws. These twin aspects are manifested as
constraints in the Hamiltonian framework. We can illustrate these remarks as
follows:

\subsubsection{ Determinism}

A trajectory, by which we mean a solution to the equations of motion, is a
function $\bar f\in F(D,R)$ at which the action is an extremum. [The
extremum is defined relative to a certain class of variations around $\bar f$%
. We will not discuss the details of these variations here.]

Suppose that $\bar f$ is a possible trajectory for a specified set of
initial conditions $d^k\bar f/dt^k\mid _{t=0},k=0,1,...,n$. Since $
\TeXButton{G}{\hat{\cal G}}$ is a gauge symmetry group, $h\bar f$ is also a
trajectory. Further, since the time dependence of $h$ is at our disposal, we
can {\it choose} $h$ such that

\begin{equation}
\left. \frac{d^k(h\overline{f})(t)}{dt^k}\right| _{t=0}=\left. \frac{d^k
\overline{f}(t)}{dt^k}\right| _{t=0},k=0,1,...,n.
\end{equation}
This does not constrain $h$ to be trivial for {\it all }time [so that we can
have $h\bar f\neq \bar f$]. The conclusion is that{\it \ there are several
possible trajectories for specified initial conditions.} [We assume of
course that $\TeXButton{G}{\hat{\cal G}}$ acts nontrivially on fields.] In
this sense, the theory does not determine the future from the present if the
state of the system is given by the values of $\overline{f}$ and its
derivatives at a given time.

In the customary formulation, determinism is restored by considering only
those functions which are invariant under $\TeXButton{G}{\hat{\cal G}}$.
These gauge invariant functions and their derivatives at a given time are
then {\it defined} to constitute the observables of the theory. (Such a
definition of observables seems to have little direct bearing on whether
they are accessible to experimental observation. It is a definition which is
{\it internal} to the theory and dictated by requirements of determinism.)

In a Hamiltonian formulation with no constraints, the specification of
Cauchy data (a point of phase space) allows us to uniquely specify the
future state of the system (at least for sufficiently small times). The
existence of a gauge symmetry group for the action $S$ thus suggests that $S$
should lead to a constrained Hamiltonian dynamics. This is in fact generally
the case. An orderly way to treat constrained dynamics is due to Dirac and
Bergmann. We will explain it briefly in the next Chapter.

\subsubsection{Conservation Laws}

The infinitesimal variation of $S$ under a gauge transformation is
characterized by arbitrary functions $\epsilon _\alpha $. If $\TeXButton{G}
{\hat{\cal G}}$ is a gauge symmetry, Noether's argument shows that there is
a charge

\begin{equation}
Q=\int_{\overline{D}}\epsilon _\alpha Q_\alpha
\end{equation}
which is a constant of motion:

\begin{equation}
\frac{dQ}{dt}=0\ .
\end{equation}
Here $\overline{D}$ is a fixed time slice of $D$. Since the $\epsilon
_\alpha $'s are arbitrary functions, we can conclude that

\begin{equation}
Q_\alpha =0\ .
\end{equation}
Thus the generators of the gauge symmetry group vanish.

In electromagnetism, the analogues of (2.10) are Gauss' law

\begin{equation}
\overrightarrow{\nabla }\cdot \overrightarrow{E}+J_0=0
\end{equation}
and the vanishing of the canonical momentum $\pi ^0$ conjugate to $A_0$. The
nonabelian generalizations of these equations are well known.

In the Hamiltonian framework, the equations $Q_\alpha =0$ become first class
constraints [cf. Chapter 3]. Quantization of the system often becomes highly
nontrivial in their presence.

\subsection{The Lagrangian}

We will assume as previously that the theories we consider admit a choice of
time. The configuration space in such a theory is usually identified with $%
F( \overline{D},{\bf R}^1)$, where $\overline{D}$ is a fixed time slice of $%
D $. It is clear however that for precision, we should write $\overline{D}_t$
for the slice of $D$ at time $t$. The customary hypothesis is that $
\overline{D}_t$ for different $t$ are diffeomorphic and that there is a
natural identification of points of $\overline{D}_t$ for different times.
Under these circumstances (which we assume), we are justified in writing $
\overline{D}$.

As an example, consider a field theory on a four dimensional manifold with
the topology of Minkowski space $M^4$. Slices at different times $t$ give
different three dimensional subspaces ${\bf R}_t^3$. Without further
considerations, there is no natural identification of points of these
spaces, that is, there is as yet no obvious meaning to the identity of
spatial points for observations at different times. What is done in practice
is as follows: On $M^4$, there is an action of the time translation group $%
\{U_\tau \mid -\infty <\tau <\infty \}$. The latter maps ${\bf R}_t^3$ to $%
{\bf R}_{t+\tau }^3$ in a smooth, invertible way. We then identify all
points in ${\bf R}_t^3$ and ${\bf R}_{t+\tau }^3$ which are carried into
each other by time translations $U_{\pm \tau }$. In the conventional
coordinates $(\overrightarrow{x},t)$,

\begin{equation}
U_\tau (\overrightarrow{x},t)=(\overrightarrow{x},t+\tau )
\end{equation}
and we think of $\overrightarrow{x}$ as referring to the {\it same} three
dimensional point for all times.

A field $f\in F(D,R)$ restricted to a given time $t$ is a function on $
\overline{D}_t$. Since we have an identification of points of $\overline{D}%
_t $ for different $t$, the field $f$ can be regarded as a one dimensional
family of functions $f_t\in F(\overline{D},R)$ parametrized by time. We have
thus established a correspondence

\begin{equation}
F(D,R)\rightarrow F({\bf R}^1,F(\overline{D},R))
\end{equation}
between functions appropriate to the action principle and curves in the
configuration space $F(\overline{D},R)$.

The Lagrangian is a function of ``coordinates and velocities.'' That is, it
is a function of a point $\alpha \in F(\overline{D},R)$ on the configuration
space and of the tangent $\dot \alpha $ to this space at this point. This
new space (a point of which is a point and a tangent at that point of the
configuration space) is called the tangent bundle $T$ $F(\overline{D},R)$ on
the configuration space.

When the action is reconstructed from the Lagrangian by the formula

\begin{equation}
S=\int dt\quad L(\alpha (t),\dot \alpha (t)),
\end{equation}
we are integrating $L$ along curves in the tangent bundle. This curve is not
arbitrary since we require that $\dot \alpha (t)=d\alpha (t)/dt$. Such a
curve in the tangent bundle is the ``lift of a curve'' from the
configuration space. (It is defined by a ``second order'' vector field in
the tangent bundle). With this restriction on curves, a curve in the tangent
bundle is uniquely determined by a curve in $F(\overline{D},R)$. Since such
a curve in turn defines a function in $F(D,R)$, we recover the original
interpretation of the action as a function on $F(D,R)$.

We need to investigate the action of the gauge group on the tangent bundle.
It turns out that in its action on the tangent bundle, the gauge group, in
its simplest version, is associated to the global group

\begin{equation}
\underline{G}\;\TeXButton{s}{\;s\;\llap{$\bigcirc$}}\;G=\{(\ell ,h)\mid \ell
\in \underline{G},\quad h\in G\}
\end{equation}
where $G$ is the global group appropriate for $\TeXButton{G}{\hat{\cal G}}$,
\underline{$G$} is its Lie algebra and the group multiplication is

\begin{equation}
(\ell ^{\prime },h^{\prime })(\ell ,h)=(\ell ^{\prime }+Ad\ h^{\prime }\
\ell ,h^{\prime }h)
\end{equation}
[The sense in which the gauge group appropriate for the Lagrangian formalism
can be thought of as associated with (2.15) will be explained below.] Here $%
Ad\ h^{\prime }$ is the adjoint action of $h^{\prime }$ on \underline{$G$}.
In the notation common in physics literature,

\begin{equation}
Ad\ h^{\prime }\ \ell =h^{\prime }\ell \ h^{\prime -1}.
\end{equation}
Thus \underline{$G$}\ $\TeXButton{s}{\;s\;\llap{$\bigcirc$}}\;G$ is the
semi-direct product of \underline{$G$} with $G$. This result has been
discussed before by Sudarshan and Mukunda.

We denote the gauge group associated to $\TeXButton{G}{\hat{\cal G}}$ at a
given time by ${\cal G}$. It consists of functions $F(\overline{D},G)=\{h\}$
with group multiplication defined by

\begin{equation}
(hh^{\prime })(\overline{d})=h(\bar d)h^{\prime }(\overline{d}),\quad
\overline{d}\in \overline{D}.
\end{equation}
The Lie algebra \underline{$G$} is a group under addition and its associated
gauge group $F(\overline{D},\underline{G})$ at a given time will be denoted
by \underline{${\cal G}$}. Finally the gauge group associated to $\underline{%
G}\;\TeXButton{s}{\;s\;\llap{$\bigcirc$}}\;G$ at a given time will be
denoted by ${\cal \underline{G}\;\;}\TeXButton{s}{\;s\;\llap{$\bigcirc$}}\;%
{\cal G}$.

In contrast to elements of ${\cal G}$, elements of the group $\TeXButton{G}
{\hat{\cal G}}$ introduced earlier had arbitrary time dependence. These two
groups are to be carefully distinguished although both have been called
gauge groups.

The group law (2.16) can be established by examining the way the action of
the gauge group $\TeXButton{G}{\hat{\cal G}}$ ``projects down'' to an action
on coordinates and velocities. A function $f\in F(D,R)$ is transformed to $%
hf $. Thus the curve $\{\alpha (t)\in F(\overline{D},{\bf R}^1)\}\ $($t$
being time) is transformed into $\{(h\alpha )(t)\}$ where $h(t)\in {\cal G}$
is time dependent. Thus a point of the tangent bundle is transformed
according to

\begin{equation}
(\alpha (t),\frac{d\alpha (t)}{dt}=\dot \alpha (t))\rightarrow (h(t)\alpha
(t),h(t)\frac{d\alpha (t)}{dt}+\ell (t)h(t)\alpha (t))
\end{equation}
where $l(t)\equiv \frac{dh(t)}{dt}h(t)^{-1}\in $\underline{${\cal G}$}. In
(2.19), all time dependences can henceforth be ignored since we are
examining the action of the gauge group restricted to $TF(\overline{D},R)$
at a given time. In writing (2.19), we have also assumed that the action of
the gauge group is local in time, that is that

\begin{equation}
(h\alpha )(t)=h(t)\alpha (t)\ .
\end{equation}
If $(h\alpha )(t)$ depends on $h(t)$ as well as (say) its derivatives $%
d^kh(t)/dt^k$, (2.19) will have to be modified. For Yang-Mills theories,
this actually happens. (See below). \ We prefer to illustrate the idea
without this complication. With this assumption, we can write

\begin{equation}
(\ell ,h)\in \underline{{\cal G}}\;\TeXButton{s}{\;s\;\llap{$\bigcirc$}}\;%
{\cal G},\quad (\ell ,h)(\alpha ,\dot \alpha )=(h\alpha ,h\dot \alpha +\ell
(h\alpha ))\ .
\end{equation}

The group multiplication (2.16) follows from

\begin{equation}
\begin{array}{c}
(\ell ^{\prime },h^{\prime })(h\alpha ,h\dot \alpha +\ell (h\alpha
))=(h^{\prime }h\alpha ,h^{\prime }h\dot \alpha +(h^{\prime }\ell h^{\prime
-1})(h^{\prime }h\alpha )+\ell ^{\prime }(h^{\prime }h\alpha )) \\
=(h^{\prime }h\alpha ,h^{\prime }h\dot \alpha +(\ell ^{\prime }+Adh^{\prime
}\ell )(h^{\prime }h\alpha ))= \\
(\ell ^{\prime }+Adh^{\prime }\ell ,h^{\prime }h)(\alpha ,\dot \alpha )
\end{array}
\end{equation}

The preceding considerations are easily illustrated by Yang-Mills theory
where the vector potential $A_\mu $ has values in the Lie algebra \underline{%
$G$} of the global group $G$ and transforms as follows:

\begin{equation}
A_\mu \rightarrow hA_\mu h^{-1}+h\partial _\mu h^{-1}.
\end{equation}
Thus at a fixed time,

\begin{equation}
(\ell ,h)A_i=hA_ih^{-1}\ ,
\end{equation}

\begin{equation}
(\ell ,h)A_0=hA_0h^{-1}-\ell
\end{equation}
where

\begin{equation}
\ell =\dot hh^{-1}\ .
\end{equation}
The group multiplication law (2.21) follows by considering the application
of $(\ell ^{\prime },h^{\prime })$ to the left hand sides of (2.24) and
(2.25).

The transformation (2.25) on the configuration space variable $A_0$ is not
local in time since (2.26) involves $dh/dt$. Nonetheless, the group
multiplication (2.21) is unaffected.

The space on which the group is supposed to act however is not the space of $%
A_\mu $, but of $(A_\mu ,\dot A_\mu )$. If we consider the subspace $%
(A_i,\dot A_i)$, since (2.24) does not involve $\dot h$, we find the group
\underline{${\cal G}$}$\;\TeXButton{s}{\;s\;\llap{$\bigcirc$}}\;{\cal G}$.
However, the argument has to be modified if $\dot A_0$ is considered since
its transformation involves $\dot \ell $. An element of the gauge group is
now a triple $(\ell ,\dot \ell ,h)$ with the action

\begin{equation}
(\ell ,\dot \ell ,h)(A_0,\dot A_0)=(hA_0h^{-1}-\ell ,h\dot A_0h^{-1}+[\ell
,hA_0h^{-1}]-\dot \ell )
\end{equation}
and the multiplication law

\begin{equation}
\left( \ell _1,\dot \ell _1,h_1)(\ell _2,\dot \ell _2,h_2)=(\ell _1+h_1\ell
_2h_1^{-1},\dot \ell _1+[\ell _1,h_1\ell _2h_1^{-1}]+h_1\dot \ell
_2h_1^{-1},h_1h_2\right) \ .
\end{equation}
The action of $(\ell ,\dot \ell ,h)$ on $(A_i,\dot A_i)$ is obtained from
taking the derivative of (2.24). In this action, $\dot \ell $ is passive.

The general gauge group ${\cal G}_L$ at the Lagrangian level can thus in
general involve $\ell $$,\dot \ell ,\ddot \ell ,...$.

The group of {\it constant} functions from $\overline{D}$ to $G$ is what is
often called the global symmetry group. Since it is isomorphic to $G$, we
can denote it by the same symbol $G$. It is a subgroup of ${\cal G}$ if all
constant functions are allowed in ${\cal G}$. Thus, if the boundary
conditions do not eliminate any such constant function, we can conclude the
following: Since observables are gauge invariant or invariant under ${\cal G}
$, they are invariant under the global group $G$. That is, all observables
are globally neutral. But note however that there are as a rule conditions
on the elements of ${\cal G}$ so that this conclusion is not always
warranted.

\subsection{ The Hamiltonian}

The Hamiltonian framework provides an algebraic formulation of the classical
theory in terms of Poisson brackets (PB's). It is the essential step in the
quantization of the classical theory.

In Chapter 3, we outline Dirac's procedure for setting up the canonical
formalism in the presence of constraints. \ Certain subtle, but important
aspects of this procedure involving the aforementioned boundary conditions
will be explained in Chapter 4 and illustrated in Chapter 5.

\section{THE DIRAC-BERGMANN\ THEORY\ OF\ CONSTRAINTS}

\subsection{Introduction}

Constraints appear in the Hamiltonian formulation of all gauge theories we
know of. We shall be applying the Dirac-Bergmann constraint theory for the
treatment of these constraints. For readers unfamiliar with the subject, we
give a very brief summary of this theory of constraints in the discussion
which follows. [See refs. 3 and 2 for reviews and applications. They also
contains further references on this subject.]

Let $M$ be the space of ``coordinates'' appropriate to a Lagrangian $L$. It
is the space $Q$ on which equations of motion give trajectories if the
Lagrangian is of the sort treated in elementary classical mechanics. More
generally, it can be different from $Q$ especially for gauge invariant
systems. We denote the points of $M$ by $m=(m_1,m_2,...)$.

Now given any manifold $M$, it is possible to associate two spaces $TM$ and $%
T^{*}M$ to $M$. The space $TM$ is called the tangent bundle over $M$. The
coordinate of a point $(m,\dot m)[\dot m=(\dot m_1,\dot m_2,...)]$ of $TM$
can be interpreted as a position and a velocity. The Lagrangian is a
function on $TM$. The space $T^{*}M$ is called the cotangent bundle over $M$%
. The coordinate of a point $(m,p)[p=(p_1,p_2,...)]$ of $T^{*}M$ can be
interpreted as a coordinate (or a ``position'') and a momentum so that in
physicists' language, $T^{*}M$ is the phase space. At each $m,\ p$ belongs
to the vector space dual to the vector space of velocities.

Poisson brackets (PB's) can be defined for any cotangent bundle $T^{*}M$. In
the notation familiar to physicists, they read

\begin{equation}
\begin{array}{c}
\{m_i,m_j\}=\{p_i,p_j\}=0\ , \\
\{m_i,p_j\}=\delta _{ij\ .}
\end{array}
\end{equation}

Now given a Lagrangian $L$, there exists a map from $TM$ to $T^{*}M$ defined
by

\begin{equation}
(m,\dot m)\rightarrow \left( m,\frac{\partial L}{\partial \dot m}(m,\dot
m\right) .
\end{equation}
If this map is globally one to one and onto, the image of $TM$ is $T^{*}M$
and we can express velocity as a function of position and momentum. This is
the case in elementary mechanics and leads to the familiar rules for the
passage from Lagrangian to Hamiltonian mechanics.

\subsection{Constraint Analysis}

It may happen, however, that the image of $TM$ under the map (3.2) is not
all of $T^{*}M$. Suppose for instance, that it is a submanifold of $T^{*}M$
defined by the equations

\begin{equation}
P_j(m,p)=0,\qquad j=1,2,...\ .
\end{equation}
Then we are dealing with a theory with constraints. The constraints $P_j$
are said to be primary.

The functions $P_j$ do not identically vanish on $T^{*}M$. Rather their
zeros define a submanifold of $T^{*}M$. A reflection of the fact that $P_j$
are not zero functions on $T^{*}M$ is that there exist functions $g$ on $%
T^{*}M$ such that $\{g,P_j\}$ do not vanish on the surface $P_j=0$. These
functions $g$ generate canonical transformations which take a point of the
surface $P_j=0$ out of this surface. It follows that it is incorrect to take
PB's of arbitrary functions with both sides of the equations $P_j=0$ and
equate them. This fact is emphasized by rewriting (3.3), replacing the
``strong'' equality signs $=$ of these equations by ``weak'' equality signs $%
\approx \ $:

\begin{equation}
P_j(m,p)\approx 0.
\end{equation}
When $P_j(m,p)$ are weakly zero, we can in general set $P_j(m,p)$ equal to
zero only after evaluating all PB's.

In the presence of constraints, the Hamiltonian can be shown to be

\begin{equation}
\begin{array}{c}
H=\dot m_j
\frac{\partial L}{\partial \dot m_j}(m,\dot m)-L(m,\dot m)+v_jP_j(m,p) \\
\\
\equiv H_0(m,p)+v_jP_j(m,p)\qquad .
\end{array}
\end{equation}
In obtaining $H_0$ from the first two terms of the first line, one can
freely use the primary constraints. The functions $v_j$ are as yet
undetermined Lagrange multipliers. Some of them may get determined later in
the analysis while the remaining ones will continue to be unknown functions
with even their time dependence arbitrary.

Consistency of dynamics requires that the primary constraints are preserved
in time. Thus we require that

\begin{equation}
\{P_m,H\}\approx 0.
\end{equation}
These equations may determine some of the $v_j$ or they may hold identically
when the constraints $P_j\approx 0$ are imposed. Yet another possibility is
that they lead to the ``secondary constraints''

\begin{equation}
P_m^{\prime }(q,p)\approx 0.
\end{equation}
The requirement $\{P_m^{\prime },H\}\approx 0$ may determine more of the
Lagrange multipliers, lead to tertiary constraints or be identically
satisfied when (3.6) and (3.7) are imposed.. We proceed in this fashion
until no more new constraints are generated.

Let us denote all the constraints one obtains in this way by

\begin{equation}
C_k\approx 0.
\end{equation}
Dirac divides these constraints into first class and second class
constraints. First class constraints $F_\alpha \approx 0$ are those for which

\begin{equation}
\{F_\alpha ,C_k\}\approx 0,\qquad \forall k.
\end{equation}
In other words, the Poisson brackets of $F_\alpha $ with $C_k$ vanish on the
surface defined by (3.8). The remaining constraints $S_a$ are defined to be
second class.

It can be shown that

\begin{equation}
\{F_\alpha ,F_\beta \}=C_{\alpha \beta }^\gamma F_\gamma ,
\end{equation}
where $C_{\alpha \beta }^\gamma (=-C_{\beta \alpha }^\gamma )$ are functions
on $T^{*}M$. The proof is as follows: Eq.(3.9) implies that $\{F_\alpha
,F_\beta \}=C_{\alpha \beta }^\gamma F_\gamma +D_{\alpha \beta }^aS_a$. But
on using the Jacobi identity

$$
\{F_\alpha ,\{F_\beta ,Sa\}\}+\{F_\beta \{Sa,F_\alpha \}\}+\{S_a,\{F_\alpha
,F_\beta \}\}=0,
$$
we find,

$$
0\approx \{S_a,\{F_\alpha ,F_\beta \}\}\approx D_{\alpha \beta
}^b\{S_a,S_b\}.
$$
In obtaining this result, we have used (3.9) which implies that $\{F_\alpha
,S_a\}$ is of the form $\sum_k\xi _{\alpha a}^kC_k$. Now as regards $S_a$,
we have the basic property

\begin{equation}
\det (\{S_a,S_b\})\neq 0
\end{equation}
on the surface $C_k\approx 0$. Thus the matrix $(\{S_a,S_b\})$ is
nonsingular on the surface $C_k\approx 0$. It then follows that $D_{\alpha
\beta }^b$ weakly vanishes, proving (3.10).

Let ${\cal C}$ be the submanifold of $T^{*}M$ defined by the constraints:

\begin{equation}
{\cal C}=\{(m,p)\mid C_k(m,p)=0\}.
\end{equation}
Then since the canonical transformations generated by $F_\alpha $ preserve
the constraints, a point of ${\cal C}$ is mapped onto another point of $%
{\cal C}$ under the canonical transformations generated by $F_\alpha $.
Since the canonical transformations generated by $S_a$ do not preserve the
constraints, such is not the case for $S_a$.

Second class constraints can be eliminated by introducing the so-called
Dirac brackets. They have the basic property that the Dirac bracket of $S_a$
with any function on $T^{*}M$ is weakly zero. We will not go into their
details having no use for them in these lectures. Instead, we shall later
follow the alternative route of finding all functions ${\cal F}$ with zero
PB's with $S_a$. So long as we work with only such functions, we can use the
constraints $S_a\approx 0$ as strong constraints $S_a=0$ and eliminate
variables using them even before taking PB's. Assuming that there are no
first class constraints, the number $N$ of functionally independent
functions ${\cal F}$ is dimension of $T^{*}M$ -- number of $S_a,N=\dim
(T^{*}M)-s$, $s$ being the range of $a$. Thus $s$ second class constraints
eliminate $s$ variables. Since $(\{S_a,S_b\})$ is nonsingular and
antisymmetric, $s$ is even. Since $\dim (T^{*}M)$ is even as well, $N$ is
even.

\subsection{Quantization Procedure}

Let us now imagine that there are only first class constraints and that $%
{\cal C}$ is defined exclusively by the zeros of $F_\alpha $. (If there are
second class constraints $S_a$ as well, they can first be eliminated in the
manner indicated above.) Dirac's prescription for the implementation of
first class constraints in quantum theory is that they be imposed as
conditions on the physically allowed states $\mid \cdot >$:

\begin{equation}
\hat F_\alpha \mid \cdot >=0.
\end{equation}
Here $\hat F_\alpha $ is the quantum operator corresponding to the classical
function $F_\alpha $.

The following may be observed in connection with (3.12). In writing it,
there is the assumption that functions on $T^{*}M$ have been realised (in
some suitable sense) as operators on a vector space.

Since the PB's between $F$'s involve only $F$'s, this prescription is
consistent (modulo factor ordering problems). That is, both sides of the
equation

\begin{equation}
[\hat F_\alpha ,\hat F_\beta ]=iC_{\alpha \beta }^\gamma \hat F_\gamma
\end{equation}
annihilate the physical states. Here the commutator brackets $[\cdot ,\cdot
] $ are obtained from the PB's using the standard prescription of Dirac. [A
similar argument shows that we cannot impose the conditions $\hat S_a\mid
\cdot >=0$ on physical states where $\hat S_a$ is the operator corresponding
to the function $S_a$.]

An observable $\TeXButton{O}{\hat{\cal O}}$ of the theory must preserve the
condition (3.13) on the physical states. Requiring that $\TeXButton{O}
{\hat{\cal O}}\mid \cdot >$ is physical if $\mid \cdot >$ is, we find, for
the set of quantum observables $\TeXButton{A}{\hat{\cal A}}$, the condition

\begin{equation}
[\TeXButton{O}{\hat{\cal O}},\hat F_\alpha ]=id_\alpha ^\gamma (\TeXButton{O}
{\hat{\cal O}})\hat F_\gamma ,\quad \TeXButton{O}{\hat{\cal O}}\in
\TeXButton{A}{\hat{\cal A}}.
\end{equation}
For classical observables ${\cal O}$, this becomes

\begin{equation}
\{{\cal O},F_\alpha \}=d_\alpha ^\gamma ({\cal O})F_\gamma .
\end{equation}
Since the right hand side is zero on ${\cal C}$, we can regard ${\cal O}$ as
a function on ${\cal C}$ which is constant on the orbits generated by $%
F_\alpha $. If we regard these orbits as generating an equivalence relation $%
\sim $ between points of ${\cal C}$, then the classical observables are
functions on the quotient of ${\cal C}$ by $\sim $. This quotient ${\cal C}%
/\sim $ may be regarded as the physical phase space. Note that if there are $%
f$ first class constraints, then the dimension $\dim [{\cal C}/\sim ]$ of
the physical phase space is $\dim (T^{*}M)-2f$, ${\cal C}$ having dimension $%
\dim (T^{*}M)-f$ and each orbit in ${\cal C}$ having dimension $f$. [Here we
assume that there is no nontrivial subgroup of the group of canonical
transformations generated by $F_\alpha $ which leaves a point of this orbit
invariant.]

An alternative method to deal with $F_\alpha $ consists in directly finding
all the classical observables ${\cal O}${\cal \ }and the corresponding
classical PB algebra ${\cal A}$ of observables. This is the algebra of
functions on ${\cal C}/\sim $. We then quantize it by replacing $\{.,.\}$ by
$-i[.,.]$ and thus find{\cal \ $\TeXButton{A}{\hat{\cal A}}$}, and then look
for a suitable representation of $\TeXButton{A}{\hat{\cal A}}$ on a Hilbert
space. In this approach, unlike in Dirac's approach, we do not first find a
vector space $V$ of vectors $\mid \cdot >$ with the property $\hat F_\alpha
\mid \cdot >=0$. Rather, we directly look for a representation of $
\TeXButton{A}{\hat{\cal A}}$.

In many examples, ${\cal C}_{\alpha \beta }^\gamma $ are constants so that $%
F_\alpha $ generate a Lie algebra over reals and are associated with a group
in a familiar manner. This group is in fact the Hamiltonian version of the
group of gauge transformations for the action. Hence one says that first
class constraints generate gauge transformations. An important fact one can
prove is that the only undetermined Lagrange multipliers in $H$ at the end
of the constraint analysis multiply first class constraints. Since $\{{\cal O%
},F_\alpha \}\approx 0$ for an observable, it follows that the time
evolution of ${\cal O}$ does not depend on these arbitrary functions. Thus a
well defined Cauchy problem can be posed on ${\cal A}$ and the time
evolution of ${\cal O}$ can be determined uniquely from suitable initial
data. The theory is therefore deterministic if we consider only ${\cal A}$.
This ceases to be the case when nonobservables are also considered since
their time evolution is influenced by the unknown Lagrange multipliers $v_j$%
. See Chapter 1 also in connection with these remarks.

Finally, we notice that there is an important symmetry structure associated
with the first class constrained surfaces in phase space, the so-called BRST
symmetry. [See ref. 2 for literature on this subject.] It is frequently used
in the quantization of gauge theories, which are typically theories with
first class constraints. We will not touch upon these considerations since
we shall have no compelling reason for using the BRST approach to
quantization.

\section{GAUGE CONSTRAINTS IN FIELD THEORIES}

\subsection{Gauss Law Generates Asymptotically Trivial Gauge Transformations}

In previous Chapters, we have outlined the physical reasons which lead to
important distinctions between gauge invariance and invariance under time
independent symmetry transformations. We have also sketched the classical
theory of constraints and its extension to the quantum domain.

In this Chapter, we look more closely at gauge constraints in field
theories. In field theories, even classical field theories, not all
functions of fields and their conjugate momenta are admissible in the
Hamiltonian formalism [3]. This is because not all functions generate well
defined canonical transformations classically. Such functions, one presumes,
are ill defined in quantum theory as well and are thus to be excluded. The
restriction of allowed phase space functions using considerations along
these lines has profound consequences for gauge field theories. It is this
restriction which leads to the possibility of QCD $\theta $-states and
fractionally charged dyons, and to the edge states of Chern-Simons dynamics.
The purpose of this Chapter is to explain this restriction and its physical
implications.

It may be remarked that there are similar constraints on functions on the
phase space ${\cal P}$ in classical mechanics as well. Thus in classical
mechanics, we almost always deal with infinitely differentiable functions on
${\cal P}$ in order that all PB's and the finite canonical transformations
obtained therefrom are well defined. The field theoretic conditions to be
found below are conditions of this kind, and are therefore to be expected.

The sort of constraints we have in mind are best illustrated by a specific
example. Let us consider the free electromagnetic field in 3+1 dimensional
Minkowski space. Let the vector potential $A_\mu $ describe this field. The
Lagrangian for this system contains no time derivative of $A_0$ so that the
momentum field $\pi _0$ conjugate to $A_0$ vanishes weakly:

\begin{equation}
\pi _0\approx 0\ .
\end{equation}
The momentum field $\pi _i$ conjugate to $A_i$ is the electric field and it
has the equal time PB

\begin{equation}
\{A_i(x),E_j(y)\}=\delta _{ij}\delta ^3(x-y)\ ,\ x^0=y^0
\end{equation}
with $A_i$. [All fields in this Section hereafter are at equal times and $x$
for example is the same as $\vec x$.] The fields $E_i$ are not all
independent, but are also subject to the Gauss law constraint

\begin{equation}
\partial _iE_i\approx 0\ .
\end{equation}

The equations (4.1) and (4.3) constitute all the constraints in this system.
They are first class, as their mutual PB is zero.

The constraint (4.1) is easy to deal with. Its PB with $A_o$ is non-zero:

\begin{equation}
\{A_0(x),\pi _0(y)\}=\delta ^3(x-y),x^0=y^0.
\end{equation}
It follows that $A_0$ is not an observable and that we can ignore it and $%
\pi _0$ as well hereafter and consider only functions of $A_i$ and $E_i$.
The latter have zero PB's with $\pi _0$ and are thus candidates for
observables.

The constraint which merits delicacy of treatment is (4.3). Let us first
rewrite it by smearing it with a `test function' $\Lambda ^\infty \ $:

\begin{equation}
\underline{g}^\infty (\Lambda ^\infty )=\int d^3x\ \Lambda ^\infty \partial
_iE_i\approx 0\ .
\end{equation}
\underline{$g$}$^\infty (\Lambda ^{(\infty )})$ is a generator of gauge
transformations on $A_i$ and $E_i$ as shown by the PB's

\begin{equation}
\begin{array}{c}
\{A_i,
\underline{g}^\infty (\Lambda ^\infty )\}=-\partial _i\Lambda ^\infty , \\
\\
\{E_i,\underline{g}^\infty (\Lambda ^\infty )\}=0.
\end{array}
\end{equation}
The underline on \underline{$g$}$^\infty $ has been put to indicate that it
is associated with the Lie algebra of the gauge group rather than with the
gauge group. The superscripts $\infty $ are to indicate certain boundary
conditions at infinity which will emerge below.

The PB's of \underline{$g$}$^\infty $ with all quantities of interest are
not well defined unless $\Lambda ^\infty $ is suitably restricted at spatial
infinity. Such a restriction does not show up in (4.6) as it involves only
the local fields $A_i$ and $E_i$. Thus, consider for example the canonical
expressions

\begin{equation}
\begin{array}{c}
J_i=\int d^3xE_j[(\vec x\times \vec \nabla )_i\delta _{jk}+\theta
(i)_{jk}]A_k\ , \\
\\
\theta (i)_{jk}=\epsilon _{ijk}
\end{array}
\end{equation}
for generators of rotations (components of angular momentum). The PB of $J_i$
with \underline{$g$}$^\infty (\Lambda ^\infty )$ can be computed by first
evaluating it with $\partial _iE_i$ and then multiplying by $\Lambda ^\infty
$ and integrating over $x_i$. Since

\begin{equation}
\{J_i,\partial \cdot E(x)\}=-\epsilon _{ijk}x_j\partial _k\ \partial \cdot
E(x)
\end{equation}
where $\partial \cdot E\equiv \partial _iE_i$, this method gives

\begin{equation}
\begin{array}{c}
\{J_i,
\underline{g}^\infty (\Lambda ^\infty )\}=\int d^3x\Lambda ^\infty
(x)\{J_i,\partial \cdot E\}(x) \\  \\
=-\int d^3x\Lambda ^\infty (x)(\vec x\times \vec \nabla )_i\partial \cdot
E(x) \\
\\
=-\int_{\mid \vec x\mid \rightarrow \infty }d\Omega \mid \vec x\mid
^2\Lambda ^\infty (x)(\vec x\times
\frac{\vec x}{\mid \vec x\mid })_i\partial \cdot E(x) \\  \\
+\int d^3x[(\vec x\times \vec \nabla )_i\Lambda ^\infty (x)]\partial \cdot
E(x) \\
\\
=\underline{g}^\infty (\vec x\times \vec \nabla )_i\Lambda ^\infty )
\end{array}
\end{equation}
where $d\Omega $ is the usual volume form on a two-sphere and $(\vec x\times
\vec \nabla )_i\Lambda ^\infty $ is the function with value $(\vec x\times
\vec \nabla )_i\Lambda ^\infty (x)$ at $x$.

We can also compute the PB $\{J_i,\underline{g}^\infty (\Lambda ^\infty )\}$
by first evaluating the PB of \underline{$g$}$^\infty (\Lambda ^\infty )$
with the integrand of $J_i\ $:

\begin{equation}
\begin{array}{c}
\{J_i,
\underline{g}^\infty (\Lambda ^\infty )\}=\{\int d^3x\ \{E_j[(\vec x\times
\vec \nabla )_i\delta _{jk}+\theta (i)_{jk}]A_k,\ \underline{g}^\infty
(\Lambda ^\infty )\} \\  \\
=-\int d^3x\quad E_j[(\vec x\times \vec \nabla )_i\delta _{jk}+\theta
(i)_{jk}]\partial _k\Lambda ^\infty \\
\\
=-\int d^3x\quad E_j\partial _j\ (\vec x\times \vec \nabla )_i\ \Lambda
^\infty \\
\\
=-\int_{\mid \vec x\mid \rightarrow \infty }d\Omega \mid \vec x\mid ^2\quad
\frac{\vec x\cdot \vec E}{\mid \vec x\mid }(\vec x\times \vec \nabla
)_i\Lambda ^\infty +\underline{g}^\infty ((\vec x\times \vec \nabla
)_i\Lambda ^\infty )\ .
\end{array}
\end{equation}

Thus the interchange of orders of integration in the evaluation of this PB
changes its value unless conditions are imposed on $\Lambda ^\infty $. [See
Chapter 5 (cf. Eq. (5.27)) or ref. 4 for another such example.] The simplest
such condition is

\begin{equation}
\Lambda ^\infty \left( x\right) \rightarrow 0\ \text{as}\mid \vec x\mid
\rightarrow \infty
\end{equation}
at some suitable rate. [We will not have to be more specific about this rate
for the purposes of these notes.]

The condition (4.11) seems reasonable for our purposes. Besides $J_i$, there
are also other functions such as momenta $P_i$ and Lorentz boosts $K_i$
which we must require to have well define PB's with \underline{$g$}$^\infty
(\Lambda ^\infty )$, and they too can lead to boundary terms containing $%
\Lambda ^\infty $ like the one in (4.10). The condition (4.11) can serve to
eliminate all these terms and to lead to well behaved PB's.

There is another way to look upon the boundary condition (4.11). Consider
the variation of \underline{$g$}$^\infty (\Lambda ^\infty )$ under a
variation $\delta E_i$ of $E_i\ $:

\begin{equation}
\delta \underline{g}^\infty (\Lambda ^\infty )=\int_{\mid \vec x\mid
\rightarrow \infty }d\Omega \mid \vec x\mid ^2\Lambda ^\infty \ \frac{\vec
x\cdot \delta \vec E}{\mid \vec x\mid }-\int d^3x\ \partial _i\Lambda
^\infty \delta E_i\ .
\end{equation}
Now a function (or ``functional'') ${\cal F}$ of a collection of fields $%
\varphi $$^{(\alpha )}$ is said to be differentiable in $\varphi ^{(\alpha
)} $ if and only if we are able to write the variation $\delta {\cal F}$ of $%
{\cal F}$ under a variation $\delta \varphi ^{(\alpha )}$ of $\varphi
^{(\alpha )}$ in the form

\begin{equation}
\delta {\cal F}=\int d^3x{\cal F}_\alpha \delta \varphi ^\alpha .
\end{equation}

If (4.13) is possible, we then define the functional derivative $\delta
{\cal F}/\delta \varphi ^\alpha $ as ${\cal F}_\alpha \ $:

\begin{equation}
\frac{\delta {\cal F}}{\delta \varphi ^{(\alpha )}(x)}={\cal F}_\alpha
[\varphi (x)],\varphi (x)=\varphi ^1(x),\varphi ^2(x),...\ .
\end{equation}
Differentiability of phase space functions in field theory is analogous to
differentiability of phase space functions in classical mechanics and is
among the simplest conditions we can impose to obtain well defined PB's.
Comparison of (4.12) and (4.13) leads to the condition (4.11) when
\underline{$g$}$^\infty (\Lambda ^\infty )$ is required to be differentiable.

Analogous considerations involving multiple PB's suggest that phase space
functions may have to be infinitely differentiable while the requirement
that they generate well defined canonical transformations can lead to more
sophisticated conditions.

It is important to remark that if for some reason we exclude functions like $%
J_i$ from consideration, then there is no reason to impose (4.11). Thus we
really must examine the collection of all functionals of possible interest
and their PB's before deciding on appropriate boundary conditions (BC's).

We will not study such difficult matters here, and will content ourselves
with the BC (4.11). Let $T^\infty $ denote the class of test functions $%
\Lambda ^\infty $ which fulfill the BC (4.11). Then the weak equality (4.5)
is thus valid only if $\Lambda ^\infty \in T^\infty $. Using the same
symbols for quantum and classical objects, it follows also that the quantum
states $\mid \cdot >$ are annihilated only by such \underline{$g$}$^\infty
(\Lambda ^\infty )$:

\begin{equation}
\underline{g}^\infty (\Lambda ^\infty )\mid \cdot >=0\Leftrightarrow \Lambda
^\infty \in T^\infty .
\end{equation}
Furthermore, as we saw in Chapter 2, the observables commute with \underline{%
$g$}$^\infty (\Lambda ^\infty )$.

The charge operator in electrodynamics is closely related to the Gauss law
operator \underline{$g$}$^\infty (\Lambda ^\infty )$. It is best discussed
after first coupling the electromagnetic field to a charged field $\psi $
with charge density $J_0$. The Gauss law and the physical state constraints
(4.5) and (4.15) are then changed to

\begin{equation}
\begin{array}{c}
\underline{g}^\infty (\Lambda ^\infty )=\int d^3x\Lambda ^\infty [\partial
_iE_i+J_0]\approx 0\ ,\quad \Lambda ^\infty \in T^\infty , \\  \\
\underline{g}^\infty (\Lambda ^\infty )\mid \cdot >=0
\end{array}
\end{equation}
while the observables now commute with this \underline{$g$}$^\infty (\Lambda
^\infty )$.

\subsection{Internal Symmetries in\ Gauge Theories}

It is convenient at this point to introduce some definitions. A general
element $e^{i\Lambda }$ of the group ${\cal G}$ of gauge transformations (at
a fixed time) in electrodynamics is a function (at a fixed time) on ${\bf R}%
^3$ with values in $U(1)\ $:

\begin{equation}
\begin{array}{c}
e^{i\Lambda }:
{\bf R}^3\rightarrow U(1), \\  \\
x\rightarrow e^{i\Lambda (x)}.
\end{array}
\end{equation}
It acts on $A_i$ and $\psi $ according to

\begin{equation}
\begin{array}{c}
A_i\rightarrow A_i+\partial _i\Lambda , \\
\\
\psi \rightarrow e^{ie\Lambda }\ \psi \ .
\end{array}
\end{equation}
We now wish to give names to several of its subgroups of particular
interest, assuming as above that the spatial slice of spacetime is ${\bf R}%
^3.$

\underline{The group ${\cal G}^c$}: The elements of ${\cal G}^c$ approach
constant values as $\mid \vec x\mid \rightarrow \infty $. If $e^{i\Lambda
^c}\in {\cal G}^c$, we thus have

\begin{equation}
\begin{array}{c}
\Lambda ^c(x)\rightarrow
\text{ constant as }\mid \vec x\mid \rightarrow \infty \ . \\
\end{array}
\end{equation}
\underline{The group ${\cal G}^\infty $}: The elements of ${\cal G}^\infty $
approach 1 as $\mid \vec x\mid \rightarrow \infty $. Because of this
boundary condition, we can identify ${\cal G}^\infty $ with the group of
maps of the three-sphere $S^3$ to $U(1)$. This sphere is the one obtained by
identifying all ``points at $\infty $'' of ${\bf R}^3$, that is by
compactifying ${\bf R}^3$ to $S^3$ by adding a ``point at $\infty $''.

\underline{The group ${\cal G}_0^\infty $}: This is the subgroup of ${\cal G}%
^\infty $ which is continuously connected to the identity. The generators of
its Lie algebra are the Gauss law constraints \underline{$g$}$^\infty
(\Lambda ^\infty )$ for all choices of $\Lambda ^\infty \in T^\infty $.

The group $G$ which is gauged in electrodynamics is $U(1)$, while it is $%
SU(3)$, in chromodynamics. All the preceding groups can be defined (in an
obvious way) for the latter as well, and indeed for any choice of a Lie
group $G$. In every case, it is easy to verify the important result that $%
{\cal G}^\infty $ is a normal subgroup of ${\cal G}$ (and hence of ${\cal G}%
^c$)\ :

\begin{equation}
{\cal G}^{(\infty )}\lhd {\cal G\ }.
\end{equation}
And of course we have the standard result

\begin{equation}
{\cal G}_0^\infty \unlhd {\cal G}^\infty .
\end{equation}

But whereas ${\cal G}^\infty $ is the same as ${\cal G}_0^\infty $ when $G$
is abelian, that is not the case when $G$ is simple. When $G$ is simple, we
have instead the important result

\begin{equation}
{\cal G}^\infty /{\cal G}_0^\infty ={\bf Z\ ,}
\end{equation}
${\bf Z}${\bf \ }being the group of integers under addition. [We assume
throughout this Chapter that the spatial manifold has the topology of ${\bf R%
}^3$.] For such a $G$, a typical element of ${\cal G}^\infty $ which is
distinct from ${\cal G}_0^\infty $ is a winding number one transformation.
Let us display such a transformation explicitly for $G=SU(2)$. If $\tau $$%
_\alpha $ are Pauli matrices, then a winding number one element of ${\cal G}%
^\infty $ is $\hat g^\infty $ where

\begin{equation}
\begin{array}{c}
\hat g^\infty (x)=e^{i\psi (r)\tau _\alpha \hat x_\alpha }, \\
\\
r=\mid \vec x\mid ,\quad \hat x_\alpha =\frac{x_\alpha }r,
\end{array}
\end{equation}
and

\begin{equation}
\psi (0)=0,\quad \psi (\infty )=2\pi .
\end{equation}
The group generated by $\hat g^\infty {\cal G}_0^\infty $ is the group ${\bf %
Z\ }$.

Note that $\hat g^\infty $ is well defined at $r=0$ because of the condition
on $\psi (0)$ and becomes $1$ at $r=\infty $, as it should being an element
of ${\cal G}^\infty $.

The expression (4.23) is identical to Skyrme's ansatz in Skyrmion physics
[2].

The generalization of (4.23) to simple Lie groups such as $G=SU(3)$ can be
constructed by looking for example at its $SU(2)$ subgroups. Thus, if $\tau
_\alpha $ in (4.23) is replaced by $\lambda _\alpha \quad (1\leq \alpha \leq
3)$ where

\begin{equation}
\lambda _\alpha =\left[
\begin{array}{ccc}
\tau _\alpha &  & 0 \\
&  & 0 \\
0 & 0 & 0
\end{array}
\right] \ ,
\end{equation}
then, for all $x$, $\hat g^\infty (x)$ is contained in a fixed $SU(2)$
subgroup of $SU(3)$ (realised as 3$\times $3 unitary matrices of determinant
1) and $\hat g^\infty {\cal G}_0^\infty $ generates ${\bf Z}$.

Now any connected Lie group is the quotient of the direct product of simple
and abelian Lie groups by discrete abelian groups (which could be trivial).
Using this fact, the preceding results can be generalized to arbitrary Lie
groups.

We turn next to the examination of these groups in the canonical formalism
and in quantum theory, limiting ourselves to $G=U(1)$ at this stage.

Closely associated to the Gauss law generator \underline{$g$}$^\infty
(\Lambda ^\infty )$ is another function obtained therefrom by partial
integration and subsequent substitution of a new test function $\xi $ for $%
\Lambda ^{(\infty )}$. We thus consider

\begin{equation}
\underline{g}(\xi )=\int d^3x[-\partial _i\xi E_i+\xi J_0]\ .
\end{equation}
It is clear that \underline{$g$}$(\xi )$ generates gauge transformations
just as \underline{$g$}$^\infty (\Lambda ^\infty )$ does:

\begin{equation}
\begin{array}{c}
\{A_i,
\underline{g}(\xi )\}=-\partial _i\xi \ , \\  \\
\{E_i,
\underline{g}(\xi )\}=0\ , \\  \\
\{\psi ,\underline{g}(\xi )\}=\xi \psi \ .
\end{array}
\end{equation}
It furthermore appears to have no problems of differentiability in $E_i$
regardless of boundary conditions on $\xi $, in contrast to what we found
with \underline{$g$}$^\infty $. Thus we seem at first sight to have
discovered the generators of ${\cal G}$.

But this conclusion is not quite correct. In electrodynamics, we encounter
electric fields $E_i$ which fall like $1/r^2$ as $r\equiv \mid \vec x\mid
\rightarrow \infty $. If there is a charge distribution of compact support
with total charge $Q$, its Coulomb field for example behaves as follows\ :

\begin{equation}
E_i=\frac Q{r^2}\frac{\hat x_i}r+0\left( \frac 1{r^3}\right) \ \text{as}\mid
\vec x\mid \rightarrow \infty ,\hat x_i=\frac{x_i}r\ .
\end{equation}
This field in a moving frame has the behavior

\begin{equation}
E_i=\frac Q{r^2}\ v_i^{(-)}(\hat x)+0\left( \frac 1{r^3}\right) \ \text{as }%
r\rightarrow \infty
\end{equation}
where $v_i^{(-)}$ is an odd function of its argument. The existence of these
fields implies that \underline{$g$}$(\xi )$ will diverge unless we constrain
$\xi $ suitably. The simplest constraint for this purpose is

\begin{equation}
\xi =\Lambda ^c.
\end{equation}
It is also what is universally assumed. It may be that there are more
general permissible conditions on $\xi $ compatible with the existence of
\underline{$g$}$(\xi )$ and with Poincar\'e invariance. We will not however
pursue this issue further here, but content ourselves with (4.29).

${\cal G}^c$ is thus a canonically implementable group and presumably can be
realised in quantum theory as well. As it acts on fields as a group of gauge
transformations, it is also an invariance group of the Hamiltonian. In
contrast, the full group ${\cal G}$ cannot be canonically implemented.

But ${\cal G}_0^\infty $ acts trivially on states and observables because of
the Gauss law constraint. It is hence only the group

\begin{equation}
{\cal G}^c/{\cal G}_0^\infty
\end{equation}
which has a nontrivial action in the theory. As it is an invariance group of
the Hamiltonian as well, we thus conclude that ${\cal G}^c/{\cal G}_0^\infty
$ is the symmetry group of electrodynamics associated to $G=U(1)$. We will
call it the internal symmetry group. [The full symmetry group is larger,
containing for instance the Poincar\'e group.]

It is important to appreciate that the internal symmetry group in a gauge
theory is a group like ${\cal G}^c/{\cal G}_0^\infty $. It is not
necessarily $G$ and may not even contain $G$. The examples below will
illustrate these points.

But for $G=U(1)$ and when the spatial slice of spacetime is ${\bf R}^3$, it
is not difficult to show that

\begin{equation}
{\cal G}^c/{\cal G}_0^\infty =U(1)\ .
\end{equation}

The proof is as follows. As mentioned previously, ${\cal G}^\infty $ and $%
{\cal G}_0^\infty $ are identical for this case. Now if two elements $%
e^{i\Lambda _j^c}\ (j=1,2)$ of ${\cal G}^c$ are both characterized by the
same boundary values of $\Lambda _j^c$ at spatial infinity, then

\begin{equation}
\left( e^{i\Lambda _1^c}\right) \left( e^{i\Lambda _2^c}\right)
^{-1}=e^{i(\Lambda _1^c-\Lambda _2^c)}
\end{equation}
approaches the value $1$ at infinity and is hence an element of ${\cal G}%
_0^\infty $. Thus a coset $e^{i\Lambda ^c}{\cal G}_0^\infty $ is entirely
fixed by the value $\left. e^{i\Lambda ^c}\right| _\infty $ of any of its
elements $e^{i\Lambda ^c}$ at infinity, this value being independent of the
choice of this element. Furthermore, the group multiplication law

\begin{equation}
\left( e^{i\Lambda _1^c}{\cal G}_0^\infty \right) \left( e^{i\Lambda _2^c}%
{\cal G}_0^\infty \right) =e^{i(\Lambda _1^c+\Lambda _2^c)}{\cal G}_0^\infty
\end{equation}
in ${\cal G}^c/{\cal G}_0^\infty $ shows that the group multiplication law
for the coset labels $\left. e^{i\Lambda ^c}\right| _\infty $ is the
standard multiplication of phases. We have thus the result (4.32).

Let \underline{$g$}$^c(\Lambda ^c)$ denote the generators of ${\cal G}^c$.
Our discussion shows that in quantum theory, when acting on states subject
to the Gauss law constraint, all that matters is the asymptotic value $%
\Lambda ^c\mid _\infty =\lambda $ of $\Lambda ^c$. So we may as well take
the function with the constant value $\lambda $ for $\Lambda ^c$ as the test
function in \underline{$g$}$^c(\Lambda ^c)$ and call it the generator $%
Q(\lambda )$ of ${\cal G}^c/{\cal G}_0^\infty $. From (4.26), we see that

\begin{equation}
Q(\lambda )=\lambda \int d^3xJ_o\ .
\end{equation}
Hence $Q(1)$ is what is called the charge $Q$ in electrodynamics.

\subsection{Nonabelian Examples}

We will here limit ourselves to brief sketches about the structure of the
symmetry group when we stray away from electrodynamics.

In chromodynamics, with ${\bf R}^3$ as the spatial slice, the discussion of
test functions like $\Lambda ^\infty $ and $\Lambda ^c$ is similar to their
discussion in electrodynamics. For example, the Gauss law generator in
chromodynamics which generalizes \underline{$g$}$^\infty (\Lambda ^\infty )$
is

\begin{equation}
\underline{g}^\infty (\Lambda ^\infty )=\int d^3x\ Tr\Lambda ^\infty D_iE_i\
,\ \Lambda ^\infty \rightarrow 0\text{ as }r\rightarrow \infty \ .
\end{equation}
Here we have not changed the notation for the generator or $\Lambda ^\infty
,D_iE_i=\partial _iE_i+[A_i,E_i]$ and $\Lambda ^\infty $, $A_i$ and $E_i$
are valued in the Lie algebra of $SU(3)$. For example, $\Lambda ^\infty
=\Lambda _\alpha ^\infty \lambda _\alpha $ where $\lambda _\alpha $ are the
Gell-Mann matrices.

The symmetry group as before is ${\cal G}^c/{\cal G}_0^\infty $. But in this
case, ${\cal G}^\infty /{\cal G}_0^\infty $ is ${\bf Z}$ instead of being
trivial. Now one can easily show, as for electrodynamics, that ${\cal G}^c/%
{\cal G}^\infty $ is G. It is also easy to show that ${\cal G}^c/{\cal G}%
^\infty $ is $({\cal G}^c/{\cal G}_0^\infty )/({\cal G}^\infty /{\cal G}%
_0^\infty )$. In other words, the symmetry group ${\cal G}^c/{\cal G}%
_0^\infty $ is an extension of $G=SU(3)$ by ${\bf Z}={\cal G}^\infty /{\cal G%
}_0^\infty $. As elements of this ${\bf Z}$ are readily seen to commute with
elements of ${\cal G}^c/{\cal G}_0^\infty $, the extension is central. The
symmetry group is thus a central extension of $SU(3)$ by ${\bf Z}$.

This extension is actually trivial:

\begin{equation}
{\cal G}^c/{\cal G}_0^\infty =SU(3)\times {\bf Z\ }.
\end{equation}
The generators of $SU(3)$ in (4.36) can be obtained from the nonabelian
analogue of (4.26) with the help of constant test functions [valued in the
Lie algebra of $SU(3)$].

The states in quantum theory can be associated with the unitary
representations of the symmetry group (4.37)\ .

The group ${\bf Z}$ has unitary irreducible representations $\rho _\theta $
which are in one-to-one correspondence with the points of the circle $S^1$.
The image of $n\in Z$ in the UIR $\rho _\theta $ is $e^{in\theta }$:

\begin{equation}
\rho _\theta :n\rightarrow e^{in\theta }.
\end{equation}
The angle $\theta $ here is the famous QCD $\theta $ parameter.

The UIR's of $SU(3)$ in (4.37) account for colour in QCD.

A more complicated and interesting example is the 't Hooft-Polyakov model
for monopoles [5]. It is a model of an $SO(3)$ or $SU(2)$ gauge theory which
in its simplest version contains a real Higgs field $\varphi =(\varphi
_1,\varphi _2,\varphi _3)$ transforming like an $SU(2)$ triplet. The vacuum
value $<\varphi >$ of $\varphi $ is a constant nonzero vector which we may
take to be $(0,0,v),v\neq 0$. The $U(1)$ or $SO(2)$ group of rotations in
the 1-2 plane leaves this $<\varphi >$ invariant so that $SO(3)$ is said to
be spontaneously broken to $U(1)$ in this model.

It was shown by 't Hooft and Polyakov that the model admits finite energy
configurations of $\varphi _i$ and the gauge potential $A_i$ with the
asymptotic conditions

\begin{equation}
\begin{array}{c}
\varphi _i(x)\rightarrow \varphi _i^\infty (x)=v\ \hat x_i\ , \\
\\
A_i\equiv A_i^\alpha \tau _\alpha \rightarrow \frac 12\ \vec \tau \cdot \hat
x\ \partial _i\ \vec \tau \cdot \hat x
\end{array}
\end{equation}
for large $r$. It was also established that these configurations provide a
field theoretic version of Dirac's magnetic monopole.

A general element of ${\cal G}$ for the 't Hooft-Polyakov model is a map

\begin{equation}
\begin{array}{c}
g:R^3\rightarrow SU(2)\ , \\
\\
x\rightarrow g(x)\ ,
\end{array}
\end{equation}
while, for the boundary conditions (4.39)\ , the analogue of ${\cal G}^c$ is
a certain group of gauge transformations which leave $\varphi ^\infty $
invariant at $\infty $. Let ${\cal G}^c$ still denote this group. It is
defined as follows. Let $g^c\in {\cal G}^c$. Then

\begin{equation}
\begin{array}{c}
g^c:
{\bf R}^3\rightarrow SU(2)\ , \\  \\
x\rightarrow g^c(x)\ ,
\end{array}
\end{equation}

\begin{equation}
\begin{array}{c}
g^c(x)
\stackunder{r\rightarrow \infty }{\rightarrow }e^{i\alpha _\infty \vec \tau
\cdot \hat x}\ , \\
\end{array}
\end{equation}
$\alpha _\infty $ being a constant independent of $\hat x$. Such a $g_c$
clearly leaves $\varphi ^\infty $ invariant.

Next suppose that $g_j^c\ (j=1,2)$ have both the same asymptotic limit as $%
r\rightarrow \infty $. Then
\begin{equation}
g_1^c(x)^{-1}g_2^c(x)\stackunder{r\rightarrow \infty }{\rightarrow }{\bf 1}
\end{equation}
so that

\begin{equation}
g_1^{c-1}g_2^c\in {\cal G}^\infty
\end{equation}
where the elements of ${\cal G}^\infty $ as before go to identity at
infinity. A generic $g^c$ with the asymptotic behaviour (4.42) is therefore
given by

\begin{equation}
g^c=g_0^c\ g^\infty ,g^\infty \in {\cal G}^\infty ,
\end{equation}
$g_0^c$ being a particular solution of the condition (4.42)\ .

One such particular solution is

\begin{equation}
g_0^c(x)=e^{i\alpha (r)\vec \tau \cdot \hat x},
\end{equation}
where for $\alpha (r)$ we insert any one function with the properties

\begin{equation}
\begin{array}{c}
\alpha (0)=0, \\
\\
0\leq \alpha (\infty )\equiv \alpha _\infty <2\pi ,
\end{array}
\end{equation}
the last condition here eliminating the ambiguity in the determination of $%
\alpha _\infty $ from the asymptotic limit of $g_0^c\ $.

The symmetry group is ${\cal G}^c/{\cal G}_0^\infty $. An element of this
group is $g_0^c\ g^\infty {\cal G}_0^\infty $. Now if $h^\infty $ and $%
k^\infty $ are two elements of ${\cal G}^\infty $ with the same winding
number, then $h^\infty =k^\infty g_0^\infty $ for some $g_0^\infty \in {\cal %
G}_0^\infty $. Hence $h^\infty {\cal G}_0^\infty =k^\infty {\cal G}_0^\infty
$, so that we can choose any one typical winding number $n$ map for $%
g^\infty $. One such choice is specified by

\begin{equation}
g^\infty (x)=(\hat g^\infty (x))^n=e^{in\psi (r)\vec \tau \cdot \hat x}\ .
\end{equation}

We may thus choose $g_0^c\ g^\infty $ according to

\begin{equation}
\begin{array}{c}
g_0^c(x)g^\infty (x)=e^{i\beta (r)\vec \tau \cdot \hat x}, \\
\\
\beta (0)=0
\end{array}
\end{equation}
for insertion into the expression $g_0^c\ g^\infty \ {\cal G}_0^\infty $. In
contrast to (4.47), we here do not restrict $\beta (\infty )$. Further, as
two $\beta (r)$ with the same $\beta (\infty )$ give the same element of the
symmetry group, it suffices to consider one $\beta (r)$ for each $\beta
(\infty )$.

For each $\beta (\infty )$, we have thus an element $\gamma \ {\cal G}%
_0^\infty $ of the symmetry group with $\gamma (x)=e^{i\beta (r)\vec \tau
\cdot \hat x}$, this correspondence being onto the group. It is also
one-to-one. For suppose that the images $\gamma _j{\cal G}_0^\infty \
(j=1,2) $ of $\beta _j(\infty )$ are equal, $\gamma _j$ being defined by

\begin{equation}
\gamma _j(x)=e^{i\beta _j(r)\vec \tau \cdot \hat x}
\end{equation}
[$\beta _j(0)$ being of course zero.] Then

\begin{equation}
\gamma _1\gamma _2^{-1}\in {\cal G}_0^\infty \ .
\end{equation}
Since $\gamma _1\gamma _2^{-1}(x)=e^{i[\beta _1(r)-\beta _2(r)]\vec \tau
\cdot \hat x}$, it follows that

\begin{equation}
\beta _1(\infty )=\beta _2(\infty )\ .
\end{equation}

Thus the elements of ${\cal G}^c/{\cal G}_0^\infty $ are all uniquely
labelled by a real number $\beta (\infty )$. A formula similar to (4.34)
also shows that the group composition in ${\cal G}^c/{\cal G}_0^\infty $
induces addition as the group composition on $\beta (\infty )$.

We have thus proved the remarkable result due to Witten [5] that the
symmetry group ${\cal G}^c/{\cal G}_0^\infty $ is the additive group ${\bf R}%
^1$ of real numbers:

\begin{equation}
{\cal G}^c/{\cal G}_0^\infty ={\bf R}^1\ .
\end{equation}
This result is to be contrasted with (4.32), (4.37). In analogy to those
expressions, we might have anticipated the symmetry group here to be $%
U(1)\times {\bf Z}$. But it is not, it is rather the nontrivial central
extension ${\bf R}^1$ of $U(1)$ by ${\bf Z}$. The critical fact which leads
to this result is that the map $\gamma $ defined above becomes a winding
number one transformation when $\beta (\infty )=2\pi $. Had it instead been
an element of ${\cal G}_0^\infty $, it would have acted trivially on states.
In such a case, there would be periodicity of the elements of the symmetry
group in $\beta (\infty )$ and this group would contain $U(1)$.

There are striking physical consequences of (4.53). The charges associated
with $U(1)$ are quantized whereas those associated with ${\bf R}^1$ are not.
Therefore, there is the possibility of fractionally charged excitations
(dyons) of the 't Hooft-Polyakov monopole as first established by Witten [5].

It is to be noted that the symmetry group ${\bf R}^1$ of the monopole sector
does not contain $U(1)$ as a subgroup even though $\varphi $ was supposed to
spontaneously break $SU(2)$ to $U(1)$.

The result (4.53) is valid in the monopole sector. In the vacuum sector, the
symmetry group is $U(1)$ as one can readily show.

In Chapter 5, we will illustrate the application of some of the ideas
developed here to Chern-Simons theories.

\section{THE QUANTUM HALL EFFECT AND THE EDGE STATES OF CHERN-SIMONS THEORY}

\subsection{Introduction}

In this Chapter, we will review certain results due to Friedman, Sokoloff,
Widom and Srivastava, Fr\"ohlich and Kerler, and Fr\"ohlich and Zee [cf.
ref. 6 and citations therein] who show that the quantum Hall (QH) system is
related to the pure Chern Simons (CS) gauge theory. We then consider CS
theory on a disk, and using the methods of Chapter 4, show that there are
chiral currents of a conformal field theory at the edge of the disk. This
result is originally due to Witten. For the QH system, the existence of
these currents has been demonstrated from microscopic considerations by
Halperin.

As we set the speed of light $c$ equal to 1, the magnetic flux can be
measured in units of $h/e$.

\subsection{Chern-Simons Field Theory and the Quantum Hall System}

Let us begin our discussion by examining a QH system characterized by zero
longitudinal resistance. The conductivity tensor $\sigma $ can then be
written as

\begin{equation}
\sigma =\left(
\begin{array}{cc}
0 & \sigma _H \\
-\sigma _H & 0
\end{array}
\right) \ .
\end{equation}

In QH systems, $\sigma _H$ is quantized and is a rational multiple of $e^2/h$%
. The idea which emerges from the works mentioned above is that this fact
may have a universal explanation emerging from rational conformal field
theories.

As the longitudinal conductivity $\sigma _L$ is zero for a two-dimensional
system with $\sigma $ given by Eq. (5.1), the current density $j$ induced by
an electric field $E$ is given by

\begin{equation}
j^a(\vec x,t)=\sigma _H\epsilon ^{ab}E_b(\vec x,t)\ ;\quad a,b=1,2\ ;\
\epsilon ^{ab}=-\epsilon ^{ba},\epsilon ^{12}=1.
\end{equation}
Here $E_a=-F_{0a},F_{\mu \nu }$ being the electromagnetic field strength
tensor.

Now if $j^0$ is the charge density, then we have the continuity equation

\begin{equation}
\frac{\partial j^0}{\partial x^0}+\vec \nabla \cdot \vec j=0,\ x^0=t.
\end{equation}
Also $B$ and $E$ are related by the Maxwell's equation

\begin{equation}
\frac{\partial B}{\partial x^0}=-\epsilon ^{ab}\partial _aE_b,
\end{equation}
where $B=F_{12}$. Equations (5.2), (5.3) and (5.4) give

\begin{equation}
\sigma _H\frac{\partial B}{\partial x^0}=\frac \partial {\partial x^0}\ j^0.
\end{equation}
We thus obtain

\begin{equation}
j^0=\sigma _H(B+B_c).
\end{equation}
Here $B_c$ is an integration constant representing a time independent
background magnetic field.

Let us assume that the three-dimensional manifold $M$ has the topology of $%
R^1\times D$ with $D$ characterizing the two-dimensional space of the
sample, and $R^1$ describing time. Furthermore, let $\eta =(\eta _{\mu \nu
}) $ be any metric of Euclidean or Lorentzian signature on $M$. Then Eqs.
(5.2) and (5.6) can be extended to a generally covariant form valid for
arbitrary metrics as well as follows.

Let

\begin{equation}
J_{\alpha \beta }(x)=\mid Det\ \eta (x)\mid ^{-1/2}\epsilon _{\alpha \beta
\gamma }j^\gamma (x),\ x=\vec x,t
\end{equation}
and

\begin{equation}
j^\alpha (x)=\frac 12\mid Det\ \eta (x)\mid ^{1/2}\sigma _H\epsilon ^{\alpha
\beta \gamma }F_{\beta \gamma }(x).
\end{equation}
Here, $\alpha ,\beta ,\gamma =0,1,2$, $\epsilon _{\alpha \beta \gamma }$ is
the totally antisymmetric symbol with $\epsilon _{012}=1$ and $t=x^0$ is
time. By (5.7) and (5.8),

\begin{equation}
J_{\alpha \beta }(x)=\sigma _HF_{\alpha \beta }(x)\ .
\end{equation}
(5.9) reduces to (5.2) and (5.6) for a flat metric.

Using the language of differential forms, we can write Eqs. (5.9) and (5.7)
as

\begin{equation}
\begin{array}{c}
J=\sigma _HF, \\
\\
J=*j
\end{array}
\end{equation}
where $J=\frac 12J_{\alpha \beta }\ dx^\alpha \wedge dx^\beta $ and $*$ is
the Hodge dual. The one form $j(x)$ is defined as

\begin{equation}
j(x)=\sum_\alpha \left( \sum_\beta \eta _{\alpha \beta }(x)j^\beta
(x)\right) dx^\alpha .
\end{equation}
The continuity equation (5.3) can be written as

\begin{equation}
dJ=0
\end{equation}
where $d$ is the exterior derivative.

We shall assume that $\sigma _H$ is a constant. Equation (5.10) then gives
the Maxwell equations

\begin{equation}
dF=0\ .
\end{equation}
Here, we can write $F=dA^{\prime },A^{\prime }=A+A_c$ where $A_c$ is the
vector potential corresponding to a constant magnetic field $B_c$ (see Eq.
5.6), $A$ represents the vector potential of a fluctuation field due to
localized sources and $A^{\prime }$ the total vector potential.

Now, Eq. (5.12) implies that

\begin{equation}
J=da
\end{equation}
where $a$ is a one form. Equation (5.10) can then be written in terms of the
one forms $a$ and $A^{\prime }$ as

\begin{equation}
da=\sigma _HdA^{\prime }\ .
\end{equation}

We now note that this last equation can be obtained from an action principle
with the action $S_{CS}$ given by

\begin{equation}
S_{CS}=\frac 1{2\sigma _H}\int_M\ (a-\sigma _HA^{\prime })\wedge d(a-\sigma
_HA^{\prime })
\end{equation}
or in terms of components,

\begin{equation}
S_{CS}=\frac 1{2\sigma _H}\int_M\ \epsilon ^{\alpha \beta \gamma }(a_\alpha
-\sigma _HA_\alpha ^{\prime })\partial _\beta (a_\gamma -\sigma _HA_\gamma
^{\prime })\ d^3x.
\end{equation}
The overall normalization of $S_{CS}$ is here fixed by the requirement that
the coupling of $A_\mu ^{\prime }$ to $j^\mu $ is by the term $-j^\mu A_\mu
^{\prime }$ in the Lagrangian density.

The action $S_{CS}$ is the Chern-Simons action for the gauge field $a-\sigma
_HA^{\prime }$.

It is important to note at this step that the derivation of Eq. (5.16) from
the QH effect is valid only in the scaling limit when both length and
1/frequency scales are large. This is because although the continuity
equation (5.12) is exact, Eq. (5.2) is experimentally observed to be valid
only at large distance and time scales.

The action $S_{CS}$ can be naturally generalized to the case where there are
several independently conserved electric current densities $%
j^{(i)},i=1,...m. $ For example, for $m$ filled Landau levels, if one
neglects mixing of levels (which is a good approximation due to the large
gaps between Landau levels), each level can be treated as dynamically
independent with electric currents in each level being separately conserved.
We will not however pursue such generalizations here.

We will continue in the next Section with the exploration of the
relationship between the Quantum Hall system and the Chern-Simons theory and
we will demonstrate how the edge currents in a Quantum Hall system arise
naturally from the Chern-Simons theory. This result is first due to Witten.
We follow the approach of Balachandran et al. [4] [see also ref. 6] who
derive further results in Chern-Simons theory using this approach.

\subsection{Conformal Edge Currents}

The Lagrangians considered in this Section follow from (5.16) by setting

\begin{equation}
\bar a=(a-\sigma _HA^{\prime })[2\pi /\mid k\sigma _H\mid ]^{1/2}
\end{equation}
and calling $\bar a$ again as $a$, $k$ being $\mid k\mid (\frac{\mid \sigma
_H\mid }{\sigma _H})$. We do so in order to be consistent with the form of
the Chern-Simons Lagrangian most frequently encountered in the literature.

In this Section, we will use natural units where $\hbar $$=c=1$.

\subsubsection{The Canonical Formalism}

Let us start with a $U(1)$ Chern-Simons (CS) theory on the solid cylinder $%
D\times R^1$ ($D$ being a disk) with action given by

\begin{equation}
S=\frac k{4\pi }\int_{D\times R^1}ada,\ a=a_\mu dx^\mu ,\ ada\equiv a\wedge
da
\end{equation}
where $a_\mu $ is a real field.

The action $S$ is invariant under diffeos of the solid cylinder and does not
permit a natural choice of a time function. As time is all the same
indispensable in the canonical approach, we arbitrarily choose a time
function denoted henceforth by $x^0$. Any constant $x^0$ slice of the solid
cylinder is then the disc $D$ with coordinates $x^1$,$\ x^2$.

It is well known that the phase space of the action $S$ is described by the
equal time Poisson brackets (PB's)

\begin{equation}
\{a_i(x),a_j(y)\}=\epsilon _{ij}\frac{2\pi }k\delta ^2(x-y)\text{ for }%
i,j=1,2,\ \epsilon _{12}=-\epsilon _{21}=1
\end{equation}
and the constraint

\begin{equation}
\partial _ia_j(x)-\partial _ja_i(x)\equiv f_{ij}(x)\approx 0
\end{equation}
where $\approx $ denotes weak equality in the sense of Dirac. [Cf. Chapter
3.] All fields are evaluated at the same time $x^0$ in these equations, and
this will continue to be the case when dealing with the canonical formalism
or quantum operators in the remainder of the paper. The connection $a_0$
does not occur as a coordinate of this phase space. This is because, just as
in electrodynamics, its conjugate momentum is weakly zero and first class
and hence eliminates $a_0$ as an observable.

The constraint (5.21) is somewhat loosely stated. As emphasized in Chapter
4, it is important to formulate it more accurately by first smearing it with
a suitable class of ``test'' functions $\Lambda ^{(0)}$. Thus we write,
instead of (5.21),

\begin{equation}
g(\Lambda ^{(0)}):=\frac k{2\pi }\int_D\Lambda ^{(0)}(x)da(x)\approx 0.
\end{equation}
It remains to state the space ${\cal T}^{(0)}$ of test functions $\Lambda
^{(0)}$. For this purpose, we recall from Chapter 4 that a functional on
phase space can be relied on to generate well defined canonical
transformations only if it is differentiable. The meaning and implications
of this remark can be illustrated here by varying $g(\Lambda ^{(0)})$ with
respect to $a_\mu \ $,

\begin{equation}
\delta g(\Lambda ^{(0)})=\frac k{2\pi }\left( \int_{\partial D}\Lambda
^{(0)}\delta a-\int_Dd\Lambda ^{(0)}\delta a\right) ,
\end{equation}
$\partial D$ being the boundary of $D$. By definition, $g(\Lambda ^{(0)})$
is differentiable in $a$ only if the boundary term - the first term - in
(5.23) is zero. We do not wish to constrain the phase space by legislating $%
\delta a$ itself to be zero on $\partial D$ to achieve this goal. This is
because we have a vital interest in regarding fluctuations of $a$ on $%
\partial D$ as dynamical and hence allowing canonical transformations which
change boundary values of $a$. We are thus led to the following condition on
functions $\Lambda ^{(0)}$ in ${\cal T}^{(0)}$:

\begin{equation}
\Lambda ^{(0)}\mid _{\partial D}=0\ .
\end{equation}

It is useful to illustrate the sort of troubles we will encounter if (5.24)
is dropped. Consider

\begin{equation}
q(\Lambda )=\frac k{2\pi }\int_Dd\Lambda a\ .
\end{equation}
It is perfectly differentiable in $a$ even if the function $\Lambda $ is
nonzero on $\partial D$. It creates fluctuations

\begin{equation}
\delta a\mid _{\partial D}=d\Lambda \mid _{\partial D}
\end{equation}
of $a$ on $\partial D$ by canonical transformations. It is a function we
wish to admit in our canonical approach. Now consider its PB with $g(\Lambda
^{(0)})\ $:

\begin{equation}
\{g(\Lambda ^0),q(\Lambda )\}=\frac k{2\pi }\int d^2xd^2y\Lambda
^{(0)}(x)\epsilon ^{ij}[\partial _j\Lambda (y)]\left[ \frac \partial
{\partial x^i}\delta ^2(x-y)\right]
\end{equation}
where $\epsilon ^{ij}=\epsilon _{ij}$. This expression is quite ill defined
if

\begin{equation}
\Lambda ^{(0)}\mid _{\partial D}\neq 0\ .
\end{equation}
Thus integration on $y$ gives zero for (5.27). But if we integrate on $x$
first, treating derivatives of distributions by usual rules, one finds
instead,

\begin{equation}
-\int_Dd\Lambda ^0d\Lambda =-\int_{\partial D}\Lambda ^0d\Lambda \ .
\end{equation}
Thus consistency requires the condition (5.24).

The constraints $g(\Lambda ^{(0)})$ are first class since

\begin{equation}
\{g(\Lambda _1^{(0)}),g(\Lambda _2^{(0)})\}
\begin{array}[t]{l}
=\frac k{2\pi }\int_Dd\Lambda _1^{(0)}d\Lambda _2^{(0)} \\
\\
=\frac k{2\pi }\int_{\partial D}\Lambda _1^{(0)}d\Lambda _2^{(0)}\  \\
\\
=0\text{ for }\Lambda _1^{(0)},\Lambda _2^{(0)}\in {\cal T}^{(0)}.
\end{array}
\end{equation}

$g(\Lambda ^{(0)})$ generates the gauge transformation $a\rightarrow
a+d\Lambda ^{(0)}$ of $a$.

Next consider $q(\Lambda )$ where $\Lambda \mid _{\partial D}$ is not
necessarily zero. Since

\begin{equation}
\begin{array}{cl}
\{q(\Lambda ),g(\Lambda ^{(0)})\} & =-\frac k{2\pi }\int_D\ d\Lambda
d\Lambda ^{(0)} \\
&  \\
& =\frac k{2\pi }\int_{\partial D}\Lambda ^{(0)}d\Lambda =0\text{ for }%
\Lambda ^{(0)}\in {\cal T}^{(0)}\ ,
\end{array}
\end{equation}
they are first class or the observables of the theory. More precisely,
observables are obtained after identifying $q(\Lambda _1)$ with $q(\Lambda
_2)$ if $(\Lambda _1-\Lambda _2)\in {\cal T}^{(0)}$. For then,

\begin{equation}
q(\Lambda _1)-q(\Lambda _2)=-g(\Lambda _1-\Lambda _2)\approx 0\ .
\end{equation}
The functions $q(\Lambda )$ generate gauge transformations $a\rightarrow
a+d\Lambda $ involving $\Lambda $'s which do not necessarily vanish on $%
\partial D$.

It may be remarked that the expression for $q(\Lambda )$ is obtained from $%
g(\Lambda ^{(0)})$ after a partial integration and a subsequent substitution
of $\Lambda $ for $\Lambda ^{(0)}$. It too generates gauge transformations
like $g(\Lambda ^{(0)})$, but the test function spaces for the two are
different. The pair $q(\Lambda )$,$g(\Lambda ^{(0)})$ thus resemble the pair
\underline{$g$}$^c(\Lambda ^c),\underline{g}^\infty (\Lambda ^\infty )$ of
electrodynamics discussed in Chapter 4. The resemblance suggests that we
think of $q(\Lambda )$ as akin to the generator of a global symmetry
transformation. It is natural to do so for another reason as well\ : the
Hamiltonian is a constraint for a first order Lagrangian such as the one we
have here, and for this Hamiltonian, $q(\Lambda )$ is a constant of motion.

In quantum gravity, for asymptotically flat spatial slices, it is often the
practice to include a surface term in the Hamiltonian which would otherwise
have been a constraint and led to trivial evolution. However, we know of no
natural choice of such a surface term, except zero, for the CS theory.

The PB's of $q(\Lambda )$'s are easy to compute\ :

\begin{equation}
\{q(\Lambda _1),q(\Lambda _2)\}=\frac k{2\pi }\int_Dd\Lambda _1d\Lambda
_2=\frac k{2\pi }\int_{\partial D}\Lambda _1d\Lambda _2\ .
\end{equation}
Remembering that the observables are characterized by boundary values of
test functions, (5.33) shows that the observables generate a $U(1)$
Kac-Moody algebra localized on $\partial D$. [Literature must be consulted
for information on Kac-Moody algebras. Knowledge of these algebras is not
important for understanding this Chapter.] Note that it is a Kac-Moody
algebra for ``zero momentum'' or ``charge''. For if $\Lambda \mid _{\partial
D}$ is a constant, it can be extended as a constant function to all of $D$
and then $q(\Lambda )=0$. The central charges (given by the right hand side
of (5.33)) and hence the representation of (5.33) are different for $k>0$
and $k<0$, a fact which reflects parity violation by the action $S$.

Let $\theta \ (mod\ 2\pi )$ be the coordinate on $\partial D$ and $\phi $ a
free massless scalar field moving with speed $v$ on $\partial D$ and obeying
the equal time PB's

\begin{equation}
\{\phi (\theta ),\dot \phi (\theta ^{\prime })\}=\delta (\theta -\theta
^{\prime })\ .
\end{equation}
If $\mu _i$ are test functions on $\partial D$ and $\partial _{\pm
}=\partial _{x^0}\pm v\partial _\theta $, then

\begin{equation}
\left\{ \frac 1v\int \mu _1(\theta )\partial _{\pm }\phi (\theta ),\frac
1v\int \mu _2(\theta )\partial _{\pm }\phi (\theta )\right\} =\pm 2\int \mu
_1(\theta )d\mu _2(\theta ),
\end{equation}
the remaining PB's involving $\partial _{\pm }\varphi $ being zero. Also $%
\partial _{\pm }\partial _{\pm }\phi =0$. Thus the algebra of observables is
isomorphic to that generated by the left moving $\partial _{+}\phi $ or the
right moving $\partial _{-}\phi $.

\subsubsection{Quantization}

Our strategy for quantization relies on the observation that if

\begin{equation}
\Lambda \mid _{\partial D}(\theta )=e^{iN\theta },
\end{equation}
then the PB's (5.33) become those of creation and annihilation operators.
These latter can be identified with the similar operators of the chiral
fields $\partial _{\pm }\phi $.

Thus let $\Lambda _N$ be any function on $D$ with boundary value $%
e^{iN\theta }$:

\begin{equation}
\Lambda _N\mid _{\partial D}(\theta )=e^{iN\theta },N\in {\bf Z-\{}0{\bf \}.}
\end{equation}
[$N=0$ is excluded here in view of a remark above, $\Lambda _0\mid
_{\partial D}$ being a constant.] These $\Lambda _N$'s exist. All $q(\Lambda
_N)$ with the same $\Lambda _N\mid _{\partial D}$ are weakly equal and
define the same observable. Let $\left\langle q(\Lambda _N)\right\rangle $
be this equivalence class of weakly equal $q(\Lambda _N)$ and $q_N$ any
member thereof. [$q_N$ can also be regarded as the equivalence class
itself.] Their PB's follow from (5.33)\ :

\begin{equation}
\{q_N,q_M\}=-iNk\ \delta _{N+M,0}\ .
\end{equation}
The $q_N$'s are the CS constructions of the Fourier modes of a massless
chiral scalar field on the circle $S^1$.

We can now proceed to quantum field theory. Let ${\cal G}(\Lambda
^{(0)}),Q(\Lambda _N)$ and $Q_N$ denote the quantum operators for $g(\Lambda
^{(0)}),q(\Lambda _N)$ and $q_N$. We then impose the constraints

\begin{equation}
\left. {\cal G}(\Lambda ^{(0)})\mid \cdot \right\rangle =0
\end{equation}
on all quantum states. It is an expression of their gauge invariance.
Because of this equation, $Q(\Lambda _N)$ and $Q(\Lambda _N^{\prime })$ have
the same action on the states if $\Lambda _N$ and $\Lambda _N^{\prime }$
have the same boundary values. We can hence write

\begin{equation}
\left. Q_N\mid \cdot \right\rangle =\left. Q(\Lambda _N)\mid \cdot
\right\rangle .
\end{equation}
Here, in view of (5.38), the commutator brackets of $Q_N$ are

\begin{equation}
[Q_N,Q_M]=Nk\delta _{N+M,0}\ .
\end{equation}

Thus if $k>0\ (k<0),Q_N$ for $N>0(N<0)$ are annihilation operators (up to a
normalization) and $Q_{-N}$ creation operators. The ``vacuum'' $\mid 0>$ can
therefore be defined by

\begin{equation}
Q_N\mid 0>=0\ \text{if }Nk>0.
\end{equation}
The excitations are obtained by applying $Q_{-N}$ to the vacuum.

When the spatial slice is a disc, the observables are all given by $Q_N$ and
our quantization is complete. When it is not simply connected, however,
there are further observables associated with the holonomies of the
connection $a$ and they affect quantization. We will not examine
quantization for nonsimply connected spatial slices here.

The CS interaction does not fix the speed $v$ of the scalar field in (5.34)
and so its Hamiltonian, a point previously emphasized by Fr\"ohlich and
Kerler, and Fr\"ohlich and Zee. This is but reasonable. For if we could fix $%
v$, the Hamiltonian $H$ for $\phi $ could naturally be taken to be the one
for a free massless chiral scalar field moving with speed $v$. It could then
be used to evolve the CS observables using the correspondence between this
field and the latter. But we have seen that no natural nonzero Hamiltonian
exists for the CS system. It is thus satisfying that we cannot fix $v$ and
hence a nonzero $H$.

\subsection{The Chern-Simons Source as a Conformal Family}

{}From the physical point of view, it is of great interest to study
Chern-Simons dynamics in the presence of point sources. It is known that the
statistics and spin of particles are changed by interaction with the CS
field and that they acquire fractional statistics and spin for suitable
choices of the coupling strength. [Cf. ref 4 and references therein.]. For
this reason, Chern-Simons dynamics with sources can provide a useful means
to describe anyons. Also we saw in Section 5.2 that the abelian CS field
theory furnishes a description of the QH effect. The sources of the CS field
can therefore be thought of as quasiparticle excitations in the QH system,
giving us another reason to study these sources. One can also argue that
there are sound mathematical reasons for studying these sources, since their
spacetime history are connected to Wilson lines and Wilson lines are
important for knot theory.

As mentioned above, when a point source is immersed in the CS field, its
statistics is affected thereby. As interaction renormalizes statistics, it
must renormalize spin as well if, as some of us may conservatively desire,
CS dynamics incorporates the canonical spin-statistics connection. One
purpose of this Section is to discuss this spin renormalization using a
generalization of the canonical approach to source free quantum CS dynamics
developed in the last Section. We will see that the specific mechanism for
spin renormalization is a novel one: the configuration space of particle
mechanics is enlarged by a circle $S^1$. A point of $S^1$ can be regarded as
parametrizing a tangent direction or an orthonormal frame (although not
canonically). A spinless source thus ends up acquiring a configuration space
which is that of a two-dimensional rotor with translations added on. What
occurs in CS theory is a massless chiral (conformal) quantum field on this $%
S^1$ (and time) with ability to change its location in space and with
precisely the right spin to maintain the spin-statistics connection. The
necessity for framing the particle has been emphasized in the literature
before. The qualitative reason for the emergence of this frame is
regularization, which surrounds the particle with a tiny hole $H$ which is
eventually shrunk to a point. The CS action is then no longer for a disc $D$%
, but for $D\backslash H$, which is a disc with a hole. In contrast to $D$,
the latter has an additional boundary $\partial H$, which is the circle $S^1$
mentioned above. Just as $\partial D$, this boundary as well can be
associated with a massless chiral scalar field. The internal states of a CS
anyon for a fixed location on $D$ thus form an infinite dimensional family
of quantum states and are not described by just a single ray. This remark
was first stated by Witten and applies with equal force to the Quantum Hall
quasiparticle if described in the Chern-Simons framework. It is also
noteworthy that the CS source is not a first quantized framed particle, but
is better regarded as a ``particle'' with a first quantized position and a
second quantized frame. One intention of this Section is to explain these
striking results with hopefully transparent arguments.

Suppose that a spinless point source with coordinate $z$ is coupled to $%
A_\mu $ with coupling $eA_\mu (z(x^0))\dot z^\mu ,\ z^0=x^0$. The field
equation $\partial _1A_2-\partial _2A_1=0$ is thereby changed to

\begin{equation}
\partial _1A_2-\partial _2A_1=-\frac{2\pi e}k\delta ^2(x-z)\ .
\end{equation}
If ${\cal C}$ is a contour enclosing $z$ with positive orientation, then by
(5.43),

\begin{equation}
\oint_{{\cal C}}A=-\frac{2\pi e}k\ .
\end{equation}

On letting ${\cal C}$ shrink to a point, it now follows that $%
A(x)=A_j(x)dx^j $ has no definite limit when $x$ approaches $z$. This
singularity of $A$ demands regularization. A good way to regularize is to
punch a hole $H$ containing $z$, and eventually to shrink the hole to a
point.

Once this hole is made, the action is no longer for a disc $D$, but for $%
D\backslash H$, a disc with a hole. $D\backslash H$ has a new boundary $%
\partial H$ and it must be treated exactly like $\partial D$. The Gauss law
must accordingly be changed to

\begin{equation}
g(\Lambda ^{(1)})\approx 0
\end{equation}
where the new test function space ${\cal T}^{(1)}$ for $\Lambda ^{(1)}$ is
defined by

\begin{equation}
\Lambda ^{(1)}\mid _{\partial D}=\Lambda ^{(1)}\mid _{\partial H}=0\ .
\end{equation}
The quantum operator ${\cal G}(\Lambda ^{(1)})$ for $g(\Lambda ^{(1)})$
annihilates all the physical states.

There are now two KM algebras of the type (5.33), one each for $\partial D$
and $\partial H$. The former is defined by observables $q(\xi ^{(0)})$ with
test functions $\xi ^{(0)}$ which vanish on $\partial H$, the latter by
observables $q(\xi ^{(1)})$ with test functions $\xi ^{(1)}$ which vanish on
$\partial D$. Let us now define the KM generators for the outer and inner
boundaries as

\begin{equation}
\begin{array}{ccc}
q_N^{(0)}\equiv q(\xi _N^{(0)}), & \xi _N^{(0)}(\theta )\mid _{\partial
D}=e^{iN\theta }, & \xi _N^{(0)}\mid _{\partial H}=0\ ; \\
&  &  \\
q_N^{(1)}\equiv q(\xi _N^{(1)}), & \xi _N^{(1)}(\theta )\mid _{\partial
H}=e^{-iN\theta }, & \xi _N^{(1)}\mid _{\partial D}=0\ ,
\end{array}
\end{equation}
$\theta \ ($mod $2\pi )$ being an angular coordinate on $\partial H$. [The
coordinates $\theta $ on both $\partial D$ and $\partial H$ increase, say,
in the anticlockwise sense.] The corresponding quantum operators will be
denoted by $Q_N^{(0)}$ and $Q_N^{(1)}$. Note that the boundary conditions
exclude the choice $\xi ^\alpha =$ a constant nonzero function on $%
D\backslash H$. Hence we may not exclude $N=0$ now.

An interpretation of the observables localized on $\partial H$ is as
follows. Let $\theta \ ($mod $2\pi )$ be an angular coordinate on $D$ which
reduces to the $\theta $ coordinates we have fixed on $\partial D$ and $%
\partial H$. A typical $A$ compatible with (5.44) has a blip $-\frac{2\pi e}%
k\delta (\theta -\theta _0)d\theta $ localized on $\partial H$ at $\theta _0$%
. The behaviour of a general $A$ on $\partial H$ can be duplicated by an
appropriate superposition of these blips. The observable $q(\xi ^{(1)})$ has
zero PB with the left side of (5.44) and hence preserves the flux enclosed
by ${\cal C}$. In fact, the finite canonical transformation generated by $%
q(\xi ^{(1)})$ changes $A$ to $A+d\xi ^{(1)}$ where the fluctuation $d\xi
^{(1)}$ creates zero net flux through ${\cal C}$. All $A$ compatible with
(5.44) can be generated from any one $A$, such as an $A$ with a blip, by
these transformations. Thus the KM algebra of observables $Q_N^{(1)}$ on $%
\partial H$ generates all connections on $\partial H$ with a fixed flux from
any one of these connections.

We have now reproduced Witten's observation that the CS anyon or the CS
version of the quantum Hall quasiparticle is a conformal family.

A point of $\partial H$ can be regarded as a frame (alluded to previously)
attached to the particle. The restriction (pull back) of a connection $A$ to
$\partial H$ can be regarded as a field on these frames. It follows from
this remark that the observables localised at $\partial H$ can be regarded
as describing spin excitations.

We refer the reader to the original papers [4] for further developments of
the approach outlined here.

\section{QUANTIZATION\ AND\ MULTIPLY\ CONNECTED\ CONFIGURATION\ SPACES}

It has been mentioned in Chapter 5 that the sources coupled to the CS field
have their statistical properties changed in a way compatible with their
spin renormalisation and the spin-statistics theorem (although for reasons
of length of the article, and time available for the lectures, we have not
gone into the details of this statistics renormalisation). It is thus
natural at this point to examine the theoretical foundations of statistics
[7] and the spin-statistics theorem [8,9]. The remaining Chapters of this
review will be devoted to this task.

It has been known for some time that the statistics of identical particles
in two or more dimensions can be understood in terms of the topology of
their configuration space $Q$, their connectivity playing a particularly
significant role. In this Chapter, after having first explained why
topology, and in particular connectivity, is important for quantisation, we
will systematically develop a method of quantisation on multiply connected
spaces, providing the necessary mathematical background along the way. The
Chapter concludes with several examples of physical systems for which
multiple connectivity is significant. Chapter 7 will be our final Chapter.
There we outline a purely topological proof of the spin-statistics theorem
which completely avoids relativistic quantum field theory (RQFT) and is
entirely based on the topology of the configuration space. There are several
interesting physical systems governed by nonrelativistic dynamics such as
those of holes in a Fermi sea or excitations above that sea. The topological
proof discussed here is applicable to many of these systems whereas a RQFT
proof looks at best contrived.

Further discussion of the material of this and subsequent Chapter and
pertinent references can be found in refs. 2 and 7 to 10.

\subsection{Configuration Space and Quantum Theory}

The dynamics of a system in classical mechanics can be described by
equations of motion on a configuration space $Q$. These equations are
generally of second order in time. Thus if the position $q(t_0)$ of the
system in $Q$ and its velocity $\dot q(t_0)$ are known at some time $t_0$,
then the equations of motion uniquely determine the trajectory $q(t)$ for
all time $t$.

When the classical system is quantized, the state of a system at time $t_0$
is not specified by a position in $Q$ and a velocity. Rather, it is
described by a wave function $\psi $ which in elementary quantum mechanics
is a (normalized) function on $Q$. The correspondence between the quantum
states and wave functions however is not one to one since two wave functions
which differ by a phase describe the same state. The quantum state of a
system is thus an equivalence class $\{e^{i\alpha }\psi \mid \alpha \ $ real$%
\}$ of normalized wave functions. The physical reason for this circumstance
is that experimental observables correspond to functions like $\psi ^{*}\psi
$ which are insensitive to this phase.

In discussing the transformation properties of wave functions, it is often
convenient to enlarge the domain of definition of wave functions in
elementary quantum mechanics in such a way as to naturally describe all the
wave functions of an equivalence class. Thus instead of considering wave
functions as functions on $Q$, we can regard them as functions on a larger
space $\hat Q=Q\times S^1\equiv \{(q,e^{i\alpha })\}$. The space $\hat Q$ is
obtained by associating circles $S^1$ to each point of $Q$ and is said to be
a $U(1)$ bundle on $Q$. Wave functions on $\hat Q$ are not completely
general functions on $\hat Q$, rather they are functions with the property $%
\psi (q,e^{i(\alpha +\theta )})=\psi (q,e^{i\alpha })e^{i\theta }$. [Here we
can also replace $e^{i\theta }$ by $e^{in\theta }$ where $n$ is a fixed
integer]. Because of this property, experimental observables like $\psi
^{*}\psi $ are independent of the extra phase and are functions on $Q$ as
they should be. The standard elementary treatment which deals with functions
on $Q$ is recovered by restricting the wave functions to a surface $%
\{(q,e^{i\alpha _0})\mid q\in Q\}$ in $Q$ where $\alpha _0$ has a fixed
value. Such a choice $\alpha _0$ of $\alpha $ corresponds to a phase
convention in the elementary approach.

When the topology of $Q$ is nontrivial, it is often possible to associate
circles $S^1$ to each point of $Q$ so that the resultant space $\hat
Q=\{\hat q\}$ is not $Q\times S^1$, although there is still an action of $%
U(1)$ on $\hat Q$. We shall indicate this action by $\hat q\rightarrow \hat
qe^{i\theta }$. It is the analogue of the transformation $(q,e^{i\alpha
})\rightarrow (q,e^{i\alpha }e^{i\theta })$ we encountered earlier. We shall
require this action to be free, which means that $\hat qe^{i\theta }=\hat q$
if and only if $e^{i\theta }$ is the identity of $U(1)$. When $\hat Q\neq
Q\times S^1$, the $U(1)$ bundle $\hat Q$ over $Q$ is said to be twisted. It
is possible to contemplate wave functions which are functions on $\hat Q$
even when this bundle is twisted provided they satisfy the constraint $\psi
(\hat qe^{i\theta })=\psi (\hat q)e^{in\theta }$ for some fixed integer $n$.
If this constraint is satisfied, experimental observables being invariant
under the $U(1)$ action are functions on $Q$ as we require. However, when
the bundle is twisted, it does not admit globally valid coordinates of the
form $(q,e^{i\alpha })$ so that it is not possible (modulo certain technical
qualifications) to make a global phase choice, as we did earlier. In other
words, it is not possible to regard wave functions as functions on $Q$ when $%
\hat Q$ is twisted.

The classical Lagrangian $L$ often contains complete information on the
nature of the bundle $\hat Q$. We can regard the classical Lagrangian as a
function on the tangent bundle $T\hat Q$ of $\hat Q$. The space $T\hat Q$ is
the space of positions in $\hat Q$ and the associated velocities. When $\hat
Q$ is trivial, it is possible to reduce any such Lagrangian to a Lagrangian
on the space $TQ$ of positions and velocities associated with $Q$ thereby
obtaining the familiar description. On the other hand, when $\hat Q$ is
twisted, such a reduction is in general impossible. \ Since the equations of
motion deal with trajectories on $Q$ and not on $\hat Q$, it is necessary
that there is some principle which renders the additional $U(1)$ degree of
freedom in such a Lagrangian nondynamical. This principle is the principle
of gauge invariance for the gauged group $U(1)$. Thus under the gauge
transformation $\hat q(t)\rightarrow \hat q(t)e^{i\theta (t)}$, these
Lagrangians change by constant times $d\theta /dt$, where $t$ is time. Since
the equations of motion therefore involve only gauge invariant quantities
which can be regarded as functions of positions and velocities associated
with $Q$, these equations describe dynamics on $Q$. The Lagrangians we often
deal with split into two terms $L_0$ and $L_{WZ}$, where $L_0$ is gauge
invariant while $L_{WZ}$ changes as indicated above. This term $L_{WZ}$ has
a geometrical interpretation. It is the one which is associated with the
nature of the bundle $\hat Q$.

In particle physics, such a topological term was first discovered by\ Wess
and Zumino in their investigation of nonabelian anomalies in gauge theories.
The importance and remarkable properties of such ``Wess-Zumino terms'' have
been forcefully brought to the attention of particle physicists in recent
years because of the realization that they play a critical role in creating
fermionic states in a theory with bosonic fields and in determining the
anomaly structure of effective field theories.

In point particle mechanics, the existence and significance of Wess-Zumino
terms have long been understood. For example, such terms play an essential
role in the program of geometric quantization and related investigations
which study the Hamiltonian or Lagrangian description of particles of fixed
spin. A similar term occurs in the description of the charge-monopole system
and has also been discussed in the literature. Such terms have been found in
dual string models as well.

The Wess-Zumino term affects the equations of motion and has significant
dynamical consequences already at the classical level. Its impact however is
most dramatic in quantum theory where, as was indicated above, it affects
the structure of the state space. \ For example, in the $SU(3)$ chiral
model, it is this term which is responsible for the fermionic nature of the
Skyrmion.

The preceding remarks on the nature of wave functions in quantum theory can
be generalized by replacing the group $U(1)$ by more general abelian or
nonabelian groups. A particularly important class of physical systems where
such groups are discrete are those with multiply connected configuration
spaces.

Multiply connected configuration spaces play an important role in many
branches of physics. Examples are molecular physics, condensed matter and
quantum field theories, and quantum gravity. Exotic statistics, which has
recently assumed an important physical role in condensed matter theory for
example, can be understood in terms of the multiple connectivity of the
configuration space. In this Chapter, we will also give a few examples of
such physical systems.

As a prelude to the discussion of multiply connected configuration spaces,
we shall first generalize the preceding remarks on the nature of wave
functions.

The arguments above which led to the consideration of $U(1)$ bundles on $Q$
were based on the observation that since only observables like $\psi
^{*}\psi $ are required to be functions on $Q$, it is permissible to
consider wave functions $\psi $ which are functions on a $U(1)$ bundle $\hat
Q$ over $Q$ provided all wave functions fulfill the property $\psi (\hat
qe^{i\theta })=\psi (\hat q)e^{in\theta }$. We shall now show that we can
meet this requirements on observables even with vector valued wave functions
$\psi =(\psi _1,...,\psi _K)$ which are functions on an $H$ bundle $\bar Q$
over $Q$, the group $H$ not being necessarily $U(1)$.

The general definition of an $H$ bundle $\bar Q$ over $Q$ is as follows. In
an $H$ bundle $\bar Q=\{\bar q\}$ over $Q$, there is an action $\bar
q\rightarrow \bar qh$ of the group $H=\{h\}$ on $\bar Q$ with the property

\begin{equation}
\bar q=\bar qh\ \text{if and only if\ }h=\ \text{identity }e.
\end{equation}
As indicated earlier, such an action of a group $H$ is said to be free.
Furthermore, in an $H$ bundle, when all points of $\bar Q$ connected by this
$H$ action are identified, we get back the space $Q$. The space $Q$ is thus
the quotient of $\bar Q$ by the $H$ action:

\begin{equation}
Q=\bar Q/H.
\end{equation}
A point of $Q$ can be thought of the set of all points $\{\bar qh\mid h\in
H\}\equiv \bar qH$ connected to $\bar q$ by the $H$ action.

If the action of $H$ on $\bar Q$ is written in the form $\bar q\rightarrow
\rho (h^{-1})\bar q\equiv \bar qh$, then $\rho (h_1)\rho (h_2)=\rho (h_1h_2)$%
. Hence the map $\rho $ : $h\rightarrow \rho (h)$ from $H$ to these
transformations on $Q$ is a homomorphism. It is in fact an isomorphism in
view of (6.1). Note that the image of $h$ under $\rho $ acts on $\bar q$
according to $\bar q\rightarrow \bar qh^{-1}$ and not according to $\bar
q\rightarrow \bar qh$. Nevertheless, following the convention in the
mathematical literature, we shall often regard the action of $h$ on $\bar Q$
as being given by $\bar q\rightarrow \bar qh$.

An example of $\bar Q$ is the trivial $H$ bundle $\bar Q=Q\times
H=\{(q,s)\mid s\in H\}$. It carries the free $H$ action $(q,s)\rightarrow
(q,s)h\equiv (q,sh)$. The quotient of $\bar Q$ by this action is $Q$. A
point of $Q$ is $q$ which can be identified with $(q,s)H$ (for any $s$).

In the mathematical literature, the space $\bar Q$ is known as the {\it %
bundle space} and $H$ is known as the {\it structure group}. The map

\begin{equation}
\begin{array}{c}
\pi :\bar Q\rightarrow Q, \\
\\
\bar q\rightarrow \bar qH
\end{array}
\end{equation}
is known as the {\it projection map}. The set of points in $\bar Q$ which
project to the same point $q$ of $Q$ under $\pi $ is known as the fibre over
$q$. The entire structure $(\bar Q,\pi ,Q,H)$ is known as a {\it principal
fibre} {\it bundle}. We shall however call $\bar Q$ itself as a principal
fibre bundle (or as an $H$ bundle).

It follows from the relation (6.2) between $\bar Q$ and $Q$ that any
function $\sigma $ on $\bar Q$ which is invariant under the $H$ action [$%
\sigma (\bar qh)=\sigma (\bar q)$ for all $h\in H$] can be regarded as a
function on $Q$. Let $h\rightarrow D(h)$ define a representation $\Gamma $
of $H=\{h\}$ by $K\times K$ unitary matrices. Let us demand of our wave
functions that they transform by $\Gamma $ under the action of $H$:

\begin{equation}
\psi _i(\bar qh)=\psi _j(\bar q)D_{ji}(h).
\end{equation}
Then for any two wave functions $\psi $ and $\psi ^{\prime }$, the expression

\begin{equation}
<\psi ,\psi ^{\prime }>(\bar q)\equiv \psi _i^{*}(\bar q)\psi _i^{\prime
}(\bar q)
\end{equation}
is invariant under $H$ and $<\psi ,\psi ^{\prime }>$ may be thought of as a
function on $Q$. If we define the scalar product $(\psi ,\psi ^{\prime })$
on wave functions by appropriately integrating $<\psi ,\psi ^{\prime }>$
over $Q$, then it is clear that there is no obvious conceptual problem in
working with wave functions of this sort.

We shall see that such vector valued wave functions with $N\geq 2$ will
occur in the general theory of multiply connected configuration spaces if $H$
is nonabelian. When that happens, as Sorkin has proved, the space of wave
functions we have described above is too large when the dimension of $\Gamma
$ exceeds 1, even when $\Gamma $ is irreducible. The reduction of this space
to its proper size will also be described following Sorkin and will be seen
to lead to interesting consequences.

A result of particular importance we shall see later and which merits
emphasis is that the quantum theory of systems with multiply connected
configuration spaces is ambiguous, there being as many inequivalent ways of
quantizing the system as there are distinct unitary irreducible
representations (UIR's) of $\pi _1(Q)$. The angle $\theta $ which labels the
vacua in QCD, for example, can be thought as the label of the distinct UIR's
of ${\bf Z}$, ${\bf Z}$ being $\pi _1(Q)$ for such a theory. As is
well-known, the quantum theories associated with different $e^{i\theta }$
are inequivalent.

\subsection{The Universal Covering Space and the Fundamental Group}

Given any manifold such as a configuration space $Q$, it is possible to
associate another manifold $\bar Q$ to $Q$ which is simply connected. The
space $\bar Q$ is known as the universal covering space of $Q$. The group $%
\pi _1(Q)=H$ acts freely on $\bar Q$ and the quotient of $\bar Q$ by this
action is $Q$. Thus $\bar Q$ is a principal fibre bundle over $Q$ with
structure group $H$. The space $\bar Q$ plays an important role in the
construction of possible quantum theories associated with $Q$. In this
Section, we shall describe the construction of $\bar Q$. We shall also
explain the concept of the fundamental group $\pi $$_1(Q)$ of $Q$ and its
action on $\bar Q$.

We shall assume in what follows that $Q$ is path-connected, that is that if $%
q_0,q_1$ are any two points of $Q$, we can find a continuous curve $q(t)\in
Q $ with $q(0)=q_0,q(1)=q_1$.

The first step in the construction of $\bar Q$ is the construction of the
path space ${\cal P}Q$ associated with $Q$. Let $q_0$ be any point of $Q$
which once chosen is held fixed in all subsequent considerations. Then $%
{\cal P}Q$ is the collection of all paths which start at $q_0$ and end at
any point $q$ of $Q$. We shall denote the paths ending at $q$ by $\Gamma
_q,\tilde \Gamma _q,\Gamma _q^{\prime }$ etc. It is to be noted that these
paths $\Gamma _q$ are {\em oriented} and {\em unparametrized}. The former
means that they are to be regarded as {\em starting} at the base point $q_0$
and {\em ending} at $q$. Each of these paths has thus an arrow attached
pointing from $q_0$ to $q$. The implication of the statement that $\Gamma _q$
is ``unparametrized'' is that (besides its orientation) only its
geographical location in $Q$ matters. If we introduce a parameter $s$ to
label points of $\Gamma _q$ and write the associated parametrized path as

\begin{equation}
\gamma _q=\{\gamma _q(s)\mid \gamma _q(0)=q_0,\ \gamma _q(1)=q\},
\end{equation}
then $\Gamma _q$ is the equivalence class of all such parametrized paths
(with parameters compatible with the orientation of $\Gamma _q$) with the
same location in $Q$.

We next introduce an equivalence relation $\sim $ on the paths known as
homotopy equivalence. We say that two paths $\Gamma _q$ and $\tilde \Gamma
_q $ with the same end point $q$ are {\em homotopic} and write

\begin{equation}
\Gamma _q\sim \tilde \Gamma _q
\end{equation}
if $\Gamma _q$ can be continuously deformed to $\tilde \Gamma _q$ while
holding $q$ (and of course $q_0$) fixed.

A more formal definition of homotopy equivalence is the following: If there
exists a continuous family of paths $\Gamma _q(t)\ [0\leq t\leq 1]$ in $Q$
(all from $q_0$ to $q$) such that

\begin{equation}
\Gamma _q(0)=\Gamma _q,\ \Gamma _q(1)=\tilde \Gamma _q,
\end{equation}
then $\Gamma _q\sim \tilde \Gamma _q$.

Let $[\Gamma _q]$ denote the equivalence class of all paths ending at $q$
which are homotopic to $\Gamma _q$. The {\em universal covering space} $\bar
Q$ of $Q$ is just the collection of all these equivalence classes:

\begin{equation}
\bar Q=\{[\Gamma _q]\}.
\end{equation}
It can be shown that $\bar Q$ is simply connected.

Of particular interest to us are the equivalence classes $[\Gamma _{q_0}]$
of all loops $\Gamma _{q_0}$ starting and ending at $q_0$. These equivalence
classes have a natural group structure. The group product is defined by

\begin{equation}
[\Gamma _{q_0}][\tilde \Gamma _{q_0}]=[\Gamma _{q_0}\cup \tilde \Gamma
_{q_0}]\text{,}
\end{equation}
where in the loop $\Gamma _{q_0}\cup \tilde \Gamma _{q_0}$, we first
traverse $\Gamma _{q_0}$ and then traverse $\tilde \Gamma _{q_0}$. The
inverse is defined by

\begin{equation}
[\Gamma _{q_0}]^{-1}=[\Gamma _{q_0}^{-1}],
\end{equation}
where the loop $\Gamma _{q_0}^{-1}$ has the same geographical location in $Q$
as $\Gamma _{q_0}$, but has the opposite orientation. The identity $e$ is
the equivalence class of the loop consisting of the single point $q_0$. It
is clear that

\begin{equation}
[\Gamma _{q_0}][\Gamma _{q_0}^{-1}]=[\Gamma _{q_0}^{-1}][\Gamma _{q_0}]=e\ .
\end{equation}

The group $\pi _1(Q)$ with elements $[\Gamma _{q_0}]$ and the group
structure defined above is known as the {\em fundamental group} of $Q$. If $%
\pi _1(Q)$ is nontrivial $[\pi _1(Q)\neq \{e\}]$, the space $Q$ is said to
be multiply connected. We shall see examples of multiply connected spaces in
Section 6.3. They will show in particular that $\pi _1(Q)$ can be abelian or
nonabelian. In any case, it is always discrete.

The group $\pi _1(Q)$ has a free action on $\bar Q$. It is defined by

\begin{equation}
[\Gamma _{q_0}]:[\Gamma _q]\rightarrow [\Gamma _{q_0}][\Gamma _q]\equiv
[\Gamma _{q_0}\cup \Gamma _q],
\end{equation}
where in $\Gamma _{q_0}\cup \Gamma _q$, we first traverse $\Gamma _{q_0}$
and then traverse $\Gamma _q$. It is a simple exercise to show that this
action is free.

We now claim that the quotient of $\bar Q$ by this action is $Q$, the
associated projection map $\pi :\bar Q\rightarrow Q$ being defined by

\begin{equation}
\pi :[\Gamma _q]\rightarrow \pi ([\Gamma _q])=q.
\end{equation}
This means the following: a) All the points $[\Gamma _q],[\tilde \Gamma _q]$%
,... with the same image $q$ under $\pi $ are related by $\pi _1(Q)$ action,
and b) these are the only points related by $\pi _1(Q)$ action. To show a),
let $\tilde \Gamma _q\cup \Gamma _q^{-1}$ be the loop based at $q_0$ where
we first go along $\tilde \Gamma _q$ from $q_0$ to $q$ and then return to $%
q_0$ along $\Gamma _q$ (in a sense opposite to the orientation of $\Gamma _q$%
). It is clear that

\begin{equation}
[\tilde \Gamma _q]=[\tilde \Gamma _q\cup \Gamma _q^{-1}][\Gamma _q]\ ,\
[\tilde \Gamma _q\cup \Gamma _q^{-1}]\in \pi _1(Q).
\end{equation}
This proves a). As regards b), elements of $\pi _1(Q)$ act by attaching
loops at the starting point $q_0$ of $\Gamma _q$ and hence map $[\Gamma _q]$
to some $[\tilde \Gamma _q]$. Both $[\Gamma _q]$ and $[\tilde \Gamma _q]$
project under $\pi $ to the same point $q$ of $Q$. This proves b).

We have now proved that $\bar Q$ is a principal fibre bundle over $Q$ with
structure group $\pi _1(Q)$.

\subsection{Examples of Multiply Connected Configuration Spaces}

It is appropriate at this point to give some examples of multiply connected
spaces. We will avoid examples from gauge and gravity theories for reasons
of simplicity. There are several such relevant examples and we shall pick
three.

1. Let $x_1,x_2,...,x_N$ be $N$ distinct points in the plane ${\bf R}^2$ and
let $Q$ be the complement of the set $\{x_1,x_2,...,x_N\}$ in ${\bf R}^2$:

\begin{equation}
Q={\bf R}^2\backslash \{x_1,x_2,...,x_N\}.
\end{equation}
Thus $Q$ is the plane with $N$ holes $x_1,x_2,...,x_N$. The fundamental
group $\pi _1(Q)$ of this $Q$ is of infinite order. It is nonabelian for $%
N\geq 2$. The generators of this group are constructed as follows: Let $q_0$
be any fixed point of $Q$ and let $C_M$ be any closed curve from $q_0$ to $%
q_0$ which encloses $x_M$ and none of the remaining holes. It is understood
that $C_M$ winds around $x_M$ exactly once with a particular orientation.
Let $C_M^{-1}$ be the curve with orientation opposite to $C_M$, but
otherwise the same as $C_M$. Let $[C_M]$ and $[C_M^{-1}]=[C_M]^{-1}$ be the
homotopy classes of $C_M$ and $C_M^{-1}$. Then $\pi _1(Q)$ consists of all
possible products like $[C_M][C_{M^{\prime }}][C_{M^{\prime \prime
}}]^{-1}...$ and is the free group with generators $[C_M]$. The products of
homotopy classes are defined here as in the last Section. For example, $%
[C_M][C_{M^{\prime }}]=[C_M\cup C_{M^{\prime }}]$ where $C_M\cup
C_{M^{\prime }}$ is the curve where we first trace $C_M$ and then trace $%
C_{M^{\prime }}$. For $N=1$, the group $\pi _1(Q)$ has one generator and is $%
{\bf Z}$. The relevance of this $Q$ for the treatment of the Aharonov-Bohm
effect should be evident.

2. In the collective model of nuclei, one considers nuclei with asymmetric
shapes with three distinct moments of inertia $I_i$ along the three
principal axes. There are also polyatomic molecules such as the ethylene
molecule $C_2H_4$ which can be described as such asymmetric rotors. The
configuration space $Q$ in these cases is the space of orientations of the
nucleus or the molecule. These orientations can be described by a real
symmetric $3\times 3$ matrix $T$ (the moment of inertia tensor) with three
distinct but fixed eigenvalues $I_i$. We now show that this $Q$ has a
nonabelian fundamental group.

Any $T\in Q$ can be written in the form

\begin{equation}
\begin{array}{c}
T\equiv
{\cal R}T_0{\cal R}^{-1}\ , \\  \\
T_0=\left[
\begin{array}{ccc}
I_1 &  & 0 \\
& I_2 &  \\
0 &  & I_3
\end{array}
\right] ,
\end{array}
\end{equation}
where ${\cal R}$ being in $SO(3)$ is regarded as a real orthogonal matrix of
determinant 1. Hence $Q$ is the orbit of $T_0$ under the action of $SO(3)$
given by (6.17). If ${\cal R}_i(\pi )$ is the rotation by $\pi $ around the $%
i^{th}$ axis,

\begin{equation}
\begin{array}{c}
{\cal R}_1(\pi )=\left[
\begin{array}{ccc}
1 &  & 0 \\
& -1 &  \\
0 &  & -1
\end{array}
\right] ,\ \quad {\cal R}_2(\pi )=\left[
\begin{array}{ccc}
-1 &  & 0 \\
& 1 &  \\
0 &  & -1
\end{array}
\right] \ , \\
\\
\\
{\cal R}_3(\pi )=\left[
\begin{array}{ccc}
-1 &  & 0 \\
& -1 &  \\
0 &  & 1
\end{array}
\right] \ ,
\end{array}
\end{equation}
then $T$ is invariant under the substitution ${\cal R}\rightarrow {\cal RR}%
_i(\pi )$. So $Q$ is the space of cosets of $SO(3)$ with respect to the four
element subgroup $\{1,{\cal R}_1(\pi ),{\cal R}_2(\pi ),{\cal R}_3(\pi )\}$.

It is convenient to view this coset space as the coset space $SU(2)/H$ of $%
SU(2)$ with regard to an appropriate subgroup $H$. For this purpose let us
introduce the standard homomorphism $R:SU(2)\rightarrow SO(3)$. The
definition of $R$ is

\begin{equation}
s\tau _is^{-1}=\tau _jR_{ji}(s)\ ,\ s\in SU(2),
\end{equation}
$\tau _i$ being Pauli matrices. [Here we think of $SU(2)$ concretely as the
group of $2\times 2$ unitary matrices of determinant 1.] Then we can write
any $T$ in the form

\begin{equation}
T=R(s)T_0R(s^{-1})
\end{equation}
and hence view $Q$ as the orbit of $T_0$ under $SU(2)$. Since by (6.19),

\begin{equation}
\begin{array}{c}
R(-s)=R(s), \\
\\
R(\pm si\tau _i)=R(\pm se^{i\pi \tau _i/2})=R(s){\cal R}_i(\pi ),
\end{array}
\end{equation}
the stability group $H$ of $T_0$ is the quaternion (or binary dihedral)
group $D_8^{*}:$

\begin{equation}
H=D_8^{*}=\{\pm {\bf 1},\pm i\tau _1,\pm i\tau _2,\pm i\tau _3\}.
\end{equation}
Thus

\begin{equation}
Q=SU(2)/D_8^{*}.
\end{equation}

It is well known that $SU(2)$ is simply connected $[\pi _1(SU(2))=\{e\}].$ A
consequence of this fact [which will not be proved here] is that

\begin{equation}
\pi _1(Q)=D_8^{*}.
\end{equation}
The loops in $Q$ associated with the elements of $D_8^{*}$ can be
constructed as follows. Consider a curve $\{s(t)\}$ in $SU(2)$ from identity
to $h\in D_8^{*}\ $:

\begin{equation}
s(t)\in SU(2)\ ,\ s(0)={\bf 1}\ ,\ s(1)=h.
\end{equation}
The image of this curve in $Q$ is $\{T(t)\}$ where

\begin{equation}
T(t)=R[s(t)]T_0\ R[s(t)^{-1}]\ .
\end{equation}
Since $T(0)=T(1)=T_0$, this is a loop in $Q$ based at $T_0$. Two loops $T(t)$
and $T^{\prime }(t)$ with different $s(1)\in D_8^{*}$ are not homotopic,
whereas all loops $T(t)$ and $T^{\prime }(t)$ with the same $s(1)\in D_8^{*}$
are homotopic and form a homotopy class. Such homotopy classes can be
thought of as the elements $h$ of $\pi _1(Q)$.

The relation (6.23) shows that $Q$ is the quotient of $SU(2)$ by the free
action

\begin{equation}
s\rightarrow sh\ ,\ s\in SU(2)\ ,\ h\in D_8^{*}
\end{equation}
of $D_8^{*}$. Furthermore $\pi _1(Q)=D_8^{*}$. Therefore in this example, $%
SU(2)$ as a manifold is the universal covering space of $Q$.

There are molecules with configuration spaces $Q$ such that $\pi _1(Q)$ is
any one of the finite discrete subgroups of $SU(2)$, the binary dihedral
group being just one of these possibilities. Reference [10] can be consulted
for further discussion of this fact and for citations to the literature.

3. The last example we shall give is relevant for discussing possible
statistics of particles in $k$ spatial dimensions. Consider $N$ identical
spinless particles in ${\bf R}^k$ [for $N\geq 2$] and assume first that $%
k\geq 3$. A configuration of these particles is given by the unordered set $%
[x_1,x_2,...,x_N]$ where $x_j\in {\bf R}^k$. The set must be regarded as
unordered (so that for example $[x_1,x_2,...,x_N]=[x_2,x_1,...,x_N]$)
because of the assumed indistinguishability of the particles. Let us also
assume that no two particles can occupy the same position so that $x_i\neq
x_j$ if $i\neq j$. The resultant space of these sets can be regarded as the
configuration space $Q$ of this system. It can be shown that $\pi _1(Q)$ is
identical to the permutation group $S_N$. The closed curves in $Q$
associated with the transpositions $s_{ij}\in S_N$ of two particles can be
constructed as follows. Choose the base point $q_0$ to be $%
[x_1^0,x_2^0,...,x_N^0]$. Let $\{\gamma _{ij}(t);\ 0\leq t\leq 1\}$ be the
loop in $Q$ defined by

\begin{equation}
\begin{array}{l}
\gamma
_{ij}(t)=[x_1^0,x_2^0,...,x_{i-1}^0,x_i(t),x_{i+1}^0,...,x_{j-1}^0,x_j(t),x_{j+1}^0,...,x_N^0]\ , \\
\\
x_i(0)=x_i^0\ ,\quad \quad x_i(1)=x_j^0\ , \\
\\
x_j(0)=x_j^0\quad ,\quad x_j(1)=x_i^0\ .
\end{array}
\end{equation}
$\{\gamma _{ij}(t)\}$ is a loop since the set $[x_1^0,x_2^0,...,x_N^0]$ is
unordered. The homotopy class of this loop can be identified with $s_{ij}$.

The distinct quantum theories of this system are labelled by the UIR's of $%
S_N$ and are associated with parastatistics. Special cases of these theories
describe bosons and fermions.

We can describe the configuration space of $N$ identical particles for $k=2$
as well in a similar way. The fundamental group $\pi _1(Q)$ for $k=2$
however is not $S_N$, but a very different (infinite) group known as the
braid group $B_N$. It is because $\pi _1(Q)=B_N$ for $k=2$ that remarkable
possibilities for statistics (such as fractional statistics) arise in two
spatial dimensions.

For $N=2$, it is simple to illustrate the difference between $B_2$ and $S_2$%
. The discussion also shows why fractional statistics is possible in two
dimensions. Thus consider the square $s_{12}^2$ of the transposition for two
particles. It is easy to see that it is the homotopy class of the curve
where $x_1^0$ is held fixed, say at the origin, and $x_2$ goes around it
from $x_2^0$ to $x_2^0$. For $k=2$, that is in a plane, this curve is a loop
with $x_1^0$ at its middle. It can not be shrunk to a point since $x_i\neq
x_j$ for points of $Q$. Thus $s_{12}^2\neq $ identity $e$ for $k=2$. A
similar argument shows that no power of $s_{12}$ is $e$. The group $B_2$ is
abelian and is generated by $s_{12}$. Its UIR's are given by $%
s_{12}\rightarrow e^{i\theta }$ where $\theta $ is real. All real $\theta $
are allowed since we have argued above that no power of $s_{12}$ is $e$. We
therefore have the possibility of fractional statistics which describe
neither bosons (for which $s_{12}\rightarrow 1)$ nor fermions (for which $%
s_{12}\rightarrow -1)$ for $k=2$. [The next two Sections describe how to
realise quantum theories for distinct UIR's of $\pi _1(Q).]$

Now for $k>2$, $s_{12}^2$ is still the homotopy class of a loop like the one
described above. But this loop can be shrunk to a point for $k>2$. For
example, it can be taken to be in a plane not enclosing $x_1^0$, if
necessary after first deforming it. It can then be shrunk to a point on this
plane. Thus $s_{12}^2=$ identity $e$ and the corresponding $\pi _1(Q)$ is $%
S_2={\bf Z}_2$. There are only two UIR's of $S_2$ and they are given by $%
s_{12}\rightarrow 1$ and $s_{12}\rightarrow -1$. They describe bosons and
fermions respectively.

\subsection{Quantization on Multiply Connected Configuration Spaces}

We shall now describe the general approach to quantization when the
configuration space $Q$ is multiply connected.

As indicated previously, this quantization can be carried out by introducing
a Hilbert space ${\cal H}$ of complex functions on $\bar Q$ with a suitable
scalar product and realizing the classical observables as quantum operators
on this space. Since the classical configuration space is $Q$ and not $\bar
Q $, classical observables are functions of $q\in Q$ and of their conjugate
momenta. Let us concentrate on functions of $q$. Let $\alpha (q)$ define a
function of $q$ and let $\hat \alpha $ be the corresponding quantum
operator. The definition of $\hat \alpha $ consists in specifying the
transformed function $\hat \alpha f$ for a generic function $f\in {\cal H}$.
Thus given the function $f$, we have to specify the value of $\hat \alpha f$
at every $\bar q$. This is done by the rule

\begin{equation}
(\hat \alpha f)(\bar q)=\alpha [\pi (\bar q]f(\bar q).
\end{equation}

The group $\pi _1(Q)$ acts on ${\cal H}$. Let $t$ denote a generic element
of $\pi _1(Q)$. If $\hat t$ is the operator which represents $t$ on ${\cal H}
$, and $\hat tf$ is the transform of a function $f\in {\cal H}$ by $\hat t$,
$\hat t$ is defined by specifying the function $\hat tf$ as follows:

\begin{equation}
(tf)(\bar q)\equiv f(\bar qt).
\end{equation}

Now $\hat \alpha $ commutes with $\hat t$:

\begin{equation}
\begin{array}{ccc}
(\hat \alpha \hat tf)(\bar q) & = & \alpha [\pi (\bar q)](\hat tf)(\bar q)
\\
& = & \alpha [\pi (\bar q)]f(\bar qt)\ , \\
(\hat t\hat \alpha f)(\bar q) & = & (\hat \alpha f)(\bar qt)\quad \quad \\
& = & \alpha [\pi (\bar qt)]f(\bar qt) \\
& = & \alpha [\pi (\bar q)]f(\bar qt) \\
& = & (\hat \alpha \hat tf)(\bar q).\quad
\end{array}
\end{equation}
Here we have used the fact that $\pi (\bar qt)=\pi (\bar q)$. [See (6.14)
and the remarks which follow.]

Since the operators $\hat t$ are not all multiples of the identity operator,
Schur's lemma tells us that this representation of the observables $\hat
\alpha $ on ${\cal H}$ is not irreducible. We can proceed in the following
way to reduce it to its irreducible components. Let $\Gamma _1,\Gamma _2,...$
denote the distinct irreducible representations of $\pi _1(Q)$. Let ${\cal H}%
_\beta ^\ell \ (\beta =1,2,...)$ be the subspaces of ${\cal H}$ which
transform by $\Gamma _\ell $, $\beta $ being an index to account for
multiple occurrences of $\Gamma _\ell $ in the reduction. Let us also define

\begin{equation}
{\cal H}^{(\ell )}=\bigoplus_\beta {\cal H}_\beta ^{(\ell )}.
\end{equation}
Then

\begin{equation}
{\cal H}=\bigoplus_\ell {\cal H}^{(\ell )}.
\end{equation}

Since $\hat \alpha $ commutes with $\hat t$, it can not map a vector
transforming $\Gamma _\ell $ to one transforming by $\Gamma _m\ (m\neq \ell
) $ since $\Gamma _\ell $ and $\Gamma _m$ are inequivalent. Thus

\begin{equation}
\hat \alpha {\cal H}^{(\ell )}\subset {\cal H}^{(\ell )}.
\end{equation}
In other words, we can realize our observables on any one subspace ${\cal H}%
^{(\ell )}$ and ignore the remaining subspaces. Quantization on the
subspaces ${\cal H}^{(\ell )}$ and ${\cal H}^{(m)}$ are known to be
inequivalent when $\ell \neq m$. Thus there are at least as many distinct
ways to quantize the system as the number of inequivalent irreducible
representations of $\pi _1(Q)$. It may also be shown that the representation
of the algebra of observables on any one ${\cal H}^{(\ell )}$ is irreducible
if $\pi _1(Q)$ is abelian, while some additional reduction is possible if it
is nonabelian as shown by Sorkin and as we shall see below.

Here we have not discussed how the momentum variables conjugate to the
coordinates are realized on ${\cal H}^{(\ell )}$. It can be shown that for
the problems at hand, these momentum variables can also be consistently
realized.

\subsection{Nonabelian Fundamental Groups}

Let us now consider nonabelian $\pi _1(Q)$ in more detail. Let $\gamma _\ell
\ (\ell =1,2,...)$ denote its distinct one dimensional UIR's and let $\bar
\gamma _\alpha \ (\alpha =1,...)$ denote its distinct UIR's of dimension
greater than 1. [For simplicity, we assume here that the indexing sets for
both abelian and nonabelian UIR's are countable.] The subspaces of ${\cal H}$
which carry $\gamma _\ell $ will be called $h_k^{(\ell )}$ and the subspaces
which carry $\bar \gamma _\alpha $ will be called $\bar h_\sigma ^{(\alpha
)},k$ and $\sigma $ being indices to account for multiple occurrences of a
given UIR in the reduction of ${\cal H}$. If we set

\begin{equation}
h^{(\ell )}=\bigoplus_kh_k^{(\ell )},
\end{equation}
then as in the abelian case the algebra of observables is represented
irreducibly on $h^{(\ell )}$, and the representations on different $h^{(\ell
)}$ are inequivalent. The novelty is associated with the representations on

\begin{equation}
\bar h^{(\alpha )}=\bigoplus_\sigma \bar h_\sigma ^{(\alpha )}.
\end{equation}
They are inequivalent for different $\alpha $, but they are not irreducible.
We now show this fact.

Let $e_\sigma (j)(j=1,2,...,n>1)$ be a basis for $\bar h_\sigma ^{(\alpha )}$
chosen so that they transform in the same way under $\pi _1(Q)$ for
different $\sigma $:

\begin{equation}
\hat te_\sigma (j)=e_\sigma (k)D(t)_{kj}\ .
\end{equation}
Here $t\rightarrow D(t)$ defines the representation $\bar \gamma _\alpha $.
[Since $\alpha $ can be held fixed in the ensuing discussion, an index $%
\alpha $ has not been put on the vectors $e_\sigma (j)$ or on the matrices $%
D(t)$.]

Now if $\hat L$ is any linear operator such that $\hat Le_\sigma (j)$
transforms in the same way as $e_\sigma (j)$,

\begin{equation}
\hat t\hat Le_\sigma (j)=[\hat Le_\sigma (k)]D_{kj}(t),
\end{equation}
that is if $[\hat L,\hat t]=0$, then by Schur's lemma $\hat L$ acts only on
the index $\sigma :$

\begin{equation}
\hat Le_\sigma (j)=e_\lambda (j){\cal D}_{\lambda \sigma }(\hat L).
\end{equation}
Furthermore, again by Schur's lemma, ${\cal D}(\hat L)$ is independent of $j$%
. Since $\hat \alpha $ in (6.29) shares the preceding \ property of $\hat L$%
, it follows that

\begin{equation}
\hat \alpha e_\sigma (j)=e_\lambda (j){\cal D}_{\lambda \sigma }(\hat \alpha
).
\end{equation}
It can be shown that there is a similar formula for momentum observables as
well.

Thus the subspace spanned by the vectors $e_\sigma (j)\ [\sigma =1,2,...]$
for any fixed $j$ is invariant under the action of observables. Also, since $%
{\cal D}(\hat \alpha )$ is independent of $j$, the representation of the
algebra of observables on the subspaces associated with different $j$ are
equivalent. It is thus sufficient to retain just one such subspace, the
remaining ones may be discarded. \ When we do so, we also obtain an
irreducible representation of the algebra of observables.

Further insight into the nature of this representation is gained by working
with a ``basis'' for ${\cal H}$ consisting of states localized at points of $%
Q$. These are analogous to the states $\mid \vec x>$ which are localized at
positions $\vec x$ in the standard nonrelativistic quantum mechanics of
spinless particles. But while there is only one such linearly independent
state for a given $\vec x$, we have $\dim \pi _1(Q)\ [\equiv $ dimension of $%
\pi _1(Q)]$ worth of such linearly independent states $\{\mid \bar qt>\}$
localized at $q$, because under $\pi $, $\bar qt$ projects to $q$
independently of $t$. [Here $\bar q$ is any conveniently chosen point of $%
\bar Q$ with $\pi (\bar q)=q$.] The group $\pi _1(Q)$ acts on these states
according to

\begin{equation}
\hat s\mid \bar q>=\mid \bar qs^{-1}>\ ,\ s\in \pi _1(Q).
\end{equation}
Clearly this representation of $\pi _1(Q)$ on the subspace spanned by the
vectors \\\noindent $\{\mid \bar qt>\}=\{\mid \bar qs^{-1}>\}$ (for fixed $%
\bar q$) is isomorphic to the regular representation of $\pi _1(Q)$. As is
well known, when this representation is fully reduced, each UIR occurs as
often as its dimension. Thus each $\gamma _\ell $ occurs once and is carried
by a one dimensional vector space with basis $F^{(\ell )}$ say, while each $%
\bar \gamma _\alpha $ occurs $\dim \bar \gamma _\alpha $ times and is
carried by a vector space with basis $E_\sigma ^{(\alpha )}(j)[j,\sigma
=1,2...,\dim \bar \gamma _\alpha ]$ say. The transformation law of $E_\sigma
^{(\alpha )}(j)$ under $\pi _1(Q)$ is

\begin{equation}
\hat tE_\sigma ^{(\alpha )}(j)=E_\sigma ^{(\alpha )}(k)D_{kj}(t).
\end{equation}
According to our previous argument, the reduction of the representation of
the algebra of observables is achieved by retaining only the subspace $%
V_j(q) $ spanned by the vectors $E_\sigma ^{(\alpha )}(j)$ for a fixed $j$
[and a fixed $\alpha $].

Now every nonzero vector in $V_j(q)$ is localized at $q$. Thus even after
this reduction, there are $\dim \bar \gamma _\alpha $ linearly independent
vectors localized at $q$. In nonrelativistic quantum mechanics, if the
system has internal symmetry (or quantum numbers like intrinsic spin), the
linearly independent states localized at $\vec x$ are of the form $\mid \vec
x,m>\ (m=1,2,...,k)$ where the index $m$ carries the representation of
internal symmetry. In this case, there are $k$ linearly independent vectors
localized at $\vec x$. The situation we are finding when $\pi _1(q)$ is
nonabelian has points of resemblance to this familiar quantum mechanical
situation in the sense that here as well there are many states localized at $%
q$.

It is of interest to know the physical observables $\hat O$ which mix the
indices $\sigma $ of the basis $E_\sigma ^{(\alpha )}(j)$. That is, it is of
interest to find the observables $\hat O$ with the property

\begin{equation}
\hat OE_\sigma ^{(\alpha )}(j)=E_\lambda ^{(\alpha )}(j){\cal D}_{\lambda
\sigma }(\hat O)
\end{equation}
such that their representation on $V_j(q)$ is irreducible. There is an
elegant, but local, geometrical construction for a family of such operators
which we now describe. Consider loops from $q$ to $q$, they can be divided
into homotopy classes $[C_t(q)][t\in \pi _1(Q)]$ labelled by elements of $%
\pi _1(Q)$. The class $[C_t(q)]$ consists of closed loops which are
homotopic to each other. The labels can be so chosen that $%
[C_s(q)][C_t(q)]=[C_{st}(q)]$ where the multiplication of homotopy classes
has been described in Section 6.2. [Note however that the loops $C_t(q)$ are
based at $q$ and not at the base point $q_0$ of Section 6.2.] Pick one
closed curve $C_t(q)$ from $[C_t(q)]$ and consider the operator which
parallel transports wave functions around $C_t(q)$. It can be shown that the
change of a wave function as a result of parallel transporting it around a
loop in $C_t(q)\in [C_t(q)]$ is independent of the choice of the loop in the
class $[C_t(q)]$. Thus the parallel transport operator depends only on the
homotopy class $[C_t(q)]$ and not on the choice of the closed curve in $%
[C_t(q)]$. It can hence be denoted by $\hat O_t$. These operators $\hat O_t$
can serve as the observables we are seeking.

The above description of the operators $\hat O_t$ is rather loose however
since $\hat O_t$ is defined only if the transform $\hat O_t\psi $ of a wave
function $\psi $ is defined and this involves specifying $(\hat O_t\psi
)(\bar q)$ for all $\bar q$. Hence we must associate a homotopy class $%
[C_t(q)]$ to each $t\in \pi _1(Q)$ and all $q$. This association must be
smooth in $q$ and fulfill the property $[C_s(q)][C_t(q)]=[C_{st}(q)]$.
Consider what happens if we smoothly change $[C_t(q)]$ as $q$ is taken
around a closed loop in the homotopy class $[C_s(q)],\ s\in \pi _1(Q)$. It
is then easy to convince oneself that $[C_t(q)]$ evolves into the homotopy
class $[C_{sts^{-1}}(q)]$. When $\pi _1(Q)$ is nonabelian, $[C_t(q)]$ will
not be equal to $[C_{sts^{-1}}(q)]$ for all $t$ and $s$. A consequence is
that the operators $\hat O_t$ are not all well defined when the UIR of $\pi
_1(Q)$ defining the quantum theory is nonabelian. [Nonetheless, the
representation of the algebra of observables we have described can be shown
to be irreducible.] The obstruction in defining all the operators $\hat O_t$
here is similar to the obstruction in defining the colour group in the
presence of nonabelian monopoles or the helicity group for massless
particles in higher dimensions.[See ref. 2 for references on these topics.]

It is remarkable that when $\pi _1(Q)$ is nonabelian, quantization can lead
to a multiplicity of states all localized at the same point. The
consequences of this multiplicity have not yet been sufficiently explored in
the literature.

\subsection{The Case of the Asymmetric Rotor}

We shall now briefly illustrate these ideas by the example of the asymmetric
rotor described in Section 6.3. The treatment given here is equivalent for
example to the standard treatment molecules with $D_8^{*}$ as the symmetry
group [that is, $\pi _1(Q)$] or of nuclei with three distinct moments of
inertia in the collective model approach to nuclei. See ref. 10 in
particular in this connection.

Let $\bar Q$ be the manifold of the group $SU(2)$ and let $s$ denote a point
of $\bar Q$. We regard $s$ as a $2\times 2$ unitary matrix of determinant $1$%
. Let $D_8^{*}$ be the quaternion subgroup of $SU(2)$:

\begin{equation}
D_8^{*}=\{\pm {\bf 1},\pm i\tau _i\ (i=1,2,3)\}.
\end{equation}
It has the free action

\begin{equation}
s\rightarrow sh\ ,\ h\in D_8^{*}\equiv H
\end{equation}
on $\bar Q$. If we identify all the eight points which are taken into each
other by this action, we get a space $Q$ which as we saw in Section 6.3 is
the configuration space of the asymptotic rotor.

The group $D_8^{*}$ has five inequivalent UIR's. Four of these are abelian
and may be described as follows. In one, the trivial one, all elements of $%
D_8^{*}$ are represented by the unit operator. In one of the remaining
three, $\pm {\bf 1}$ and $\pm i\tau _1$ are represented by $+1$ while $\pm
i\tau _2$ and $\pm i\tau _3$ are represented by $-1$. The two remaining one
dimensional UIR's are constructed similarly, $\pm {\bf 1}$ and $\pm i\tau _2$
being represented by $+1$ in one and $\pm {\bf 1}$ and $\pm i\tau _3$ being
represented by $+1$ in the other. As regards the two dimensional UIR, it is
the defining representation (6.44) involving Pauli matrices.

There are thus five ways of quantizing this system. We now concentrate on
the quantization method involving the two dimensional nonabelian UIR of $%
D_8^{*}$.

A basis for all functions on $SU(2)$ are the matrix elements $D_{\rho \sigma
}^j(s)[s\in SU(2)]$ of the rotation matrices. The group $D_8^{*}=\{h\}$ acts
by operators $\hat h$ on these functions according to the rule

\begin{equation}
(\hat hD_{\rho \sigma }^j)(s)=D_{\rho \sigma }^j(sh).
\end{equation}
Since

\begin{equation}
D_{\rho \sigma }^j(sh)=D_{\rho \lambda }^j(s)D_{\lambda \sigma }^j(h)
\end{equation}
and since for integer $j$, $h\rightarrow D^j(h)$ for $h\in D_8^{*}$ defines
an abelian representation of $D_8^{*}$, we can and shall restrict $j$ to
half odd integer values.

The next step is to reduce the representation $h\rightarrow D^j(h)$ into its
irreducible components. It then splits into a direct sum of the two
dimensional UIR's (6.44). [Only the two dimensional UIR's occur in this
reduction. This is because the image of $(i\tau _i)^2$ being a $2\pi $
rotation is represented by $-1$, $j$ being half an odd integer.] The basis
vectors for the vector spaces which carry such UIR's are of the form $%
e_{\rho ,m,a}^j$, $m=1,2,...,N$; $a=1,2$ where $2N$ equals $2j+1$. Under the
transformations $s\rightarrow sh$, their behavior is given by

\begin{equation}
e_{\rho ,m,a}^j(sh)=e_{\rho ,m,b}^j(s)h_{ba}.
\end{equation}

The vector space which carries the algebra of observables irreducibly is
spanned by $e_{\rho ,m,a_0}^j$ with one fixed value $a_0$ and with $j,\rho ,$
and $m$ taking on all allowed values. The vectors $e_{\rho ,m,a^{\prime }}^j$
with the remaining values $a^{\prime }$ for $a$ are to be discarded.

When the asymmetric rotor model is used to describe nuclei, $m$ can be
interpreted in terms of the third component of angular momentum in the body
fixed frame.

We have not discussed a scalar product for this vector space. A suitable
scalar product may be

\begin{equation}
(\alpha ,\beta )=\int_{SU(2)}d\mu (s)\alpha ^{*}(s)\beta (s).
\end{equation}
Here we have regarded the elements of our vector space as functions on $%
SU(2) $ and $d\mu (s)$ is the invariant measure on $SU(2)$.

In the preceding discussion, we have not referred to a Lagrangian or a
Hamiltonian. They are of course important from a dynamical point of view.
They do now however play a critical role in the construction of the vector
space for wave functions that we have outlined because this construction is
valid for a large class of Lagrangians and Hamiltonians.

\section{TOPOLOGICAL SPIN-STATISTICS THEOREMS}

In nonrelativistic quantum mechanics or relativistic quantum field theory
(RQFT) in three or more (spatial) dimensions, one encounters two sorts of
particles or localized solitonic excitations. One of these is characterized
by tensorial states, which are invariant under $2\pi $ rotation, and the
other by spinorial states, which change sign under this rotation. If we
limit ourselves to Bose and Fermi systems, the spin-statistics correlation
in three or more dimensions amounts to the assertion that the former are
bosons and the latter are fermions. Thus according to this assertion, the
change in the phase of a state under the exchange of two identical systems
of spin $S$ is $\exp [i2\pi S]$. In two dimensions, there are more general
possibilities for spin and statistics such as fractional spin and fractional
statistics. \ But here as well, the above correlation asserts that the
exchange operation is associated with the phase $\exp [i2\pi S]$ for a spin $%
S$ ``anyon'' subject to fractional statistics. [It may be emphasized here
however that the notions of spin and statistics are more fragile in two
dimensions. There the assignment of a well-defined statistics ceases to make
sense when generic, velocity-dependent forces (``magnetic fields'') are
present; and spin is subject to a similar loss of meaning. In such
situations, the spin-statistics correlation is vacuous and our discussion
will not apply.]

There are different sorts of proofs of this correlation currently available
in the literature. One class of proofs typically uses RQFT in one of its
formulations such as the one initiated by\ Wightman, or the algebraic
formulation of quantum field theory. In the Wightman framework, for example,
it is shown that tensorial fields commute and spinorial ones anticommute for
space like separations, and this result is interpreted as a proof of the
spin-statistics connection. A second approach to the spin-statistics theorem
due to Finkelstein and Rubinstein applies to solitons or ``kinks''. It is a
``topological'' proof which does not use the heavy machinery of RQFT. It
examines the fundamental group $\pi _1(Q)$ of the configuration space $Q$
appropriate for solitons and shows that $2\pi $ rotation of a soliton and
exchange of two identical solitons are the same element of $\pi _1(Q)$. This
proof in particular does not use relativistic invariance, but does use the
facts that solitons are continuous structures in field theories and that
each soliton has its antisoliton.

The spin-statistics theorem is pertinent in disciplines such as atomic
physics where relativity or field theory does not play a significant role.
It is therefore desirable to prove it for point particles in a topological
manner that would dispense with these assumptions. We may also hope that
such a proof would make sense for topological geons in quantum gravity,
where again the assumptions of flat space quantum field theory are too
restrictive. Indeed there are reasons to hope that, once we see how such a
derivation would go, we will have an important clue to the dynamical rules
governing the change of spacetime topology. A derivation of this sort will
be outlined in this Section.

References 8 and 9 can be consulted for citations to the literature on
topological spin-statistics theorems, including those discussed here.

The existence of an antiparticle is an indispensable ingredient in the
topological proofs for solitons, and will be so here as well. The concept of
antiparticle in this context can be associated with any state which on
suitable pairing with a particle state acquires the quantum numbers of the
ground state. The proof below is thus applicable to condensed matter systems
with particle-hole excitations. \ There are however many situations in low
energy physics where even such antiparticles are not available. Electron
pair production energies being several orders of magnitude larger than
typical energies in atomic physics for example, the spin-statistics
connection hence seems to provide us an example where a high energy result
has a profound influence on low energy physics.

For purposes of simplicity, we shall assume here that the particle and
antiparticle are distinct when they have spin, although this assumption can
be dispensed with. We will not use such an assumption here when the
particles are spinless.

We may at this juncture point out an important implication of the
topological spin-statistics theorems for particles moving in ${\bf R}^d\
(d\geq 2)$: They exclude ``nonabelian'' statistics. Thus according to these
theorems, paraparticles of order $2$ and more for $d\geq 3$, and particles
associated with nonabelian braid group representations for $d=2$, could not
exist in nature.

Let us first outline the proof for spinless particles with distinct
antiparticles. As discussed in Section 6.3, in one conventional approach to
statistics in particle mechanics, the configuration space $Q_M$ for $M$
identical spinless particles in ${\bf R}^d\ (d\geq 2)$ is

\begin{equation}
Q_M=\left\{ \left[ x^{(1)},x^{(2)},...,x^{(M)}\right] \left|
\begin{array}{l}
x^{(i)}\in
{\bf R}^d;x^{(i)}\neq x^{(j)}\ {\em if}\ i\neq j; \\  \\
\left[ x^{(1)},...,x^{(i)},...,x^{(j)},...,x^{(M)}\right] \\
\\
=\left[ x^{(1)},...,x^{(j)},...,x^{(i)},...,x^{(M)}\right]
\end{array}
\right. \right\} \ .
\end{equation}
The configuration space $\bar Q_N$ for $N$ spinless antiparticles is
obtained from (7.1) by replacing $M$ by $N$ and $x^{(i)}$'s by $\bar x^{(i)}$%
's. Next consider the Cartesian product

\begin{equation}
Q_{M,N}=Q_M\times \bar Q_N,\ M,N\geq 1.
\end{equation}
Define also

\begin{equation}
\begin{array}{c}
Q_{M,0}=Q_M,\ M\geq 1, \\
\\
Q_{0,N}=\bar Q_N,\ N\geq 1,
\end{array}
\end{equation}
and introduce the vacuum (``VAC'') by setting

\begin{equation}
Q_{0,0}=\{VAC\}.
\end{equation}
The final configuration space $C_K$ is obtained by imposing an equivalence
relation $\sim $ on the disjoint union

\begin{equation}
\coprod\Sb K\ fixed \\ N+K\geq 0 \\ \\ N\geq 0\endSb \ Q_{N+K,N}
\end{equation}
which makes creation and annihilation processes possible. According to this
relation, a particle and an antiparticle at the same location ``annihilate''
to VAC, and conversely they emerge from VAC by separating from an identical
location. This is illustrated in Fig. 1 and can also be expressed in
equations as follows:

\begin{equation}
\begin{array}{c}
([x];[\bar x])\sim
\text{VAC if }x=\bar x, \\  \\
([x^{(1)},...,x^{(i)},...,x^{(N+K)}];[\bar x^{(1)},...,\bar x^{(j)},...,\bar
x^{(N)}]) \\
\\
\sim ([x^{(1)},...,
\underline{x}^{(i)},...,x^{(N+K)}];[\bar x^{(1)},...,\underline{\bar x}%
^{(j)},...,\bar x^{(N)}]) \\  \\
\text{if }x^{(i)}=\bar x^{(j)}.
\end{array}
\end{equation}
Here the underlined entries are to be deleted and equations such as $\break%
([x^{(1)},x^{(2)},\underline{x}^{(3)}];[\underline{\bar x}%
])=[x^{(1)},x^{(2)}]$ are to be understood. $C_K$ is the quotient of (7.5)
by this equivalence relation. Elements of $C_K$ which are equivalence
classes containing points such as $([x^{(1)},...,x^{(N+K)}];[\bar
x^{(1)},...,\bar x^{(N)}])$ will be denoted by $[x^{(1)},...,x^{(N+K)};\bar
x^{(1)},...,\bar x^{(N)}]$. They fulfill identities which follow from (7.6).
The significance of $K$ is that the particle number (= number of
particles-number of antiparticles) for a point of $C_K$ is $K$. Note that $%
C_K$ is infinite dimensional and not a manifold.

The spin-statistics connection for spinless particles reduces to the
statement that the particles are bosons. To establish this we will show that
the exchange operation is associated with the trivial element of $\pi
_1(C_K) $. That this topological triviality of exchange does in fact entail
Bose statistics in the ordinary sense is not something we will prove here,
plausible though it is. Chapter 6 can be consulted regarding this point.

The result that particle interchange is trivial in $\pi _1(C_K)$ will be
shown in $C_2$ adopting the following conventions, the proof for any $C_K$
being similar. The homotopy parameter $t$ will increase upwards in the
figures, their horizontal sections being ${\bf R}^d$. Following Feynman, a
``particle travelling backward in $t$'' will be used to represent an
antiparticle. We will sometimes refer to $t$ as time. The base point for
homotopy will correspond to two particles located say on the 1-axis.

The curve for exchange is Fig. 2(a) whereas the trivial curve describing
static particles is Fig. 2(g). Figures (a-g) show how to deform the first to
the last of these figures thereby demonstrating the theorem. Exchanges being
the identity of $\pi _1(C_K)$, nonabelian statistics are also excluded. Note
that $p_1$ and $p_2$ [$q_1$ and $q_2$] are VAC, and superposing them as in
the passage from Fig. 2(b) to 2(c) [2(e) to 2(f)] is a legitimate activity.

If the particle and its antiparticle are not distinct, the configuration
space is

\begin{equation}
D_K=\bigcup\limits_{M=K\,\text{mod\thinspace }2}\left\{ \left[
x^{(1)},x^{(2)},...,x^{(M)}\right] \left|
\begin{array}{l}
\left[ x^{(1)},...,x^{(i)},...,x^{(j)},...,x^{(M)}\right] \\
\\
=\left[ x^{(1)},...,x^{(j)},...,x^{(i)},...,x^{(M)}\right] ; \\
\\
\left[ x^{(1)},...,x^{(i)},...,x^{(j)},...,x^{(M)}\right] \\
\\
=\left[ x^{(1)},...,
\underline{x}^{(i)},...,\underline{x}^{(j)},...,x^{(M)}\right] \  \\  \\
\text{if}\ x^{(i)}=x^{(j)}\ .
\end{array}
\right. \right\}
\end{equation}
Here $K$ is either $0$ or $1$, underlined entries are as usual to be deleted
and we employed the convention that $[x^{(1)},x^{(2)},...,x^{(M)}]:=$VAC
when $M=0$. The spin-statistics connection and the exclusion of nonabelian
statistics can be proved for $D_K$ exactly as before.

We now turn to particles with spin. We will in a well-known way account for
spin by attaching a frame to each particle. [The physical origin of these
frames is documented further in the second paper of ref. 8.] Let ${\cal F}^d$
be the set of all frames, orthonormal with respect to the Euclidean metric
on ${\bf R}^d$ and with a fixed orientation. The generalization $Q_M^{SPIN}$
of $Q_M$ to spinning particles is then

\begin{equation}
Q_M^{SPIN}=\{[(x^{(1)},F^{(1)}),...,(x^{(M)},F^{(M)})]\}
\end{equation}
where $x^{(i)}\in {\bf R}^d,F^{(i)}\in {\cal F}^d$, the elements of $%
Q_M^{SPIN}$ are invariant under permutations of the $(x^{(i)},F^{(i)})$ and
we require $x^{(i)}\neq x^{(j)}$ if $i\neq j$. The antiparticle space $\bar
Q_N^{SPIN}$ which generalizes $\bar Q_N$ is similarly obtained. Its elements
are denoted by $[(\bar x^{(1)},\bar F^{(1)}),...,(\bar x^{(N)},\bar
F^{(N)})],\bar F^{(i)}\in \TeXButton{F bar}{\bar{\cal F}}^d$ where now $
\TeXButton{F bar}{\bar{\cal F}}^d$ is the set of orthonormal frames
oppositely oriented to elements of ${\cal F}^d$. Such an orientation
reversal is suggested by the fact that the CP or CPT transform of a left
handed particle is a right handed antiparticle. It is also suggested by the
Finkelstein-Rubinstein work. The particle and antiparticle are distinct
since ${\cal F}^d\neq \TeXButton{F bar}{\bar{\cal F}}^d$.

Our final spinning particle configuration space $C_K^{SPIN}$ for particle
number $K$ is obtained from the disjoint union

\begin{equation}
\coprod\Sb K\ Fixed \\ N+K\geq 0 \\ \\ N\geq 0\endSb \ Q_{N+K,N}^{SPIN}
\end{equation}
by specifying a condition which makes annihilation and creation possible.
[The definition of $Q_{M,N}^{SPIN}$ is essentially analogous to the
definition of $Q_{M,N}$. See the second or third paper of ref. 8 for a more
precise treatment.] For this purpose, consider for simplicity a particle $i$
and an antiparticle $j$ moving towards each other along a straight line $L$
and colliding at $\xi $. Let ${\cal P}$ be the plane through $\xi $ normal
to $L$. Our central assumption is that $i$ and $j$ annihilate at $\xi $ if
and only if the antiparticle frame $\bar F^{(j)}$ approaches the reflection
of the particle frame $F^{(i)}$ in ${\cal P}$. There is a similar rule for
pair production. These assumptions are shown in Fig. 3 for $d=2$. [The axes
of the particle (antiparticle) frames are drawn in figures with single
(double) lines]. They imply equations such as

\begin{equation}
\begin{array}{c}
\lim _{x^{(i)},\bar x^{(j)}\rightarrow \xi }\ [(x^{(i)},F^{(i)});(\bar
x^{(j)},\bar F^{(j)})]=VAC; \\
\\
\lim _{x^{(i)},\bar x^{(j)}\rightarrow \xi }\
[(x^{(1)},F^{(1)}),...,(x^{(i)},F^{(i)}),...;(\bar x^{(1)},\bar
F^{(1)},...,(\bar x^{(j)},\bar F^{(j)}),...] \\
\\
=[(x^{(1)},F^{(1)}),...,\underline{(x^{(i)},F^{(i)})},...,;(\bar
x^{(1)},\bar F^{(1)}),...,\underline{(\bar x^{(j)},\bar F^{(j)})},...]
\end{array}
\end{equation}
where the limit is taken with $x^{(i)}$ and $\bar x^{(j)}$ approaching $\xi $
along $L$ and the antiparticle frame approaching the appropriate reflection
of the particle frame (explained above) in the limit. The rest of the new
notation follows the earlier one.

The exchange diagram Fig. 2(a) is as before homotopic to\ Fig. 2(e) where
now an appropriate frame is supposed to be attached to each point of these
figures. We now show that the left hand side Fig. 4(a) of Fig. 2(e) is
homotopic to Fig. 4(b,c) where the frame of the outgoing particle undergoes $%
2\pi $ rotation as $t$ evolves, thereby showing the theorem.

The homotopy of Fig. 4(b) to Fig. 4(c) is obtained by coalescing $C$ and $D$%
. We must thus prove the homotopy of Fig. 4(a) and Fig. 4(b). For this
purpose, it is convenient to assume that the particles and antiparticle in
these pictures are moving along the 1-axis except within the dashed circle
when the particle created by pair production takes a little excursion in the
1-2 plane and then returns to the 1-axis.

The process in Fig. 4(a) is redrawn in Fig. 5, which shows only the first
two axes of the frames. At times $t<t_1$, a particle, call it $1$, is moving
to the right on 1-axis. A pair is produced at $t=t_1$, with the particle $2$
of the pair to the left of antiparticle $\bar 2$. As $t$ evolves, $1$ and $%
\bar 2$ annihilate at $t=t_2$ while 2 moves to the left on 1-axis, makes a
detour in the 1-2 plane and then returns to the 1-axis. Fig. 4(b) is
likewise redrawn in Fig. 6. \ Note that the alignment of the frames in Figs.
5 and 6 is consistent with (7.10).

A comparison of these figures shows that the left-right order of the $2-\bar
2$ pair at the moment of production is reversed in going from Fig. 5 to Fig.
6. The homotopy of Fig. 5 to\ Fig. 6 thus involves gradually changing the
production angle $\theta $ of 2 from $\pi $ as in Fig. 5 to zero as in Fig.
6. [We assume that 2 is produced in the 1-2 plane in the successive stages
of the homotopy.] Fig. 7 shows the frame of 2 as $\theta $ is so changed,
the $\bar 2$ frame being held fixed. Clearly, because of the mirror rule
involved in (7.10), the frame of $2$ rotates by $2(\theta _1-\pi )$ when $%
\theta $ decreases from $\pi $ to $\theta _1$. This means that when Fig. 5
is deformed so that $\theta $ becomes $\theta _1$, the frame of $2$ will
rotate by $2(\pi -\theta _1)$ before 2 reaches its final destination. \ This
is shown in Fig. 8. This rotation being $2\pi $ for $\theta =0$, the
homotopy of Figs. 4(a,b) is thus established.

Nonabelian statistics can be shown to be excluded by a simple extension of
the preceding arguments. Thus consider $M$ particles in $C_K^{SPIN}$ say. By
the above, the exchange $\sigma _{ij}$ of particles $i$ and $j$ is equal to $%
2\pi $ rotation $R_{2\pi }^{(i)}$ of the frame $i$. Repeating the argument,
we have further $R_{2\pi }^{(i)}=\sigma _{1i}=R_{2\pi }^{(1)}$ whence all
exchanges and all rotations are homotopic to each other. This shows that all
exchanges commute thereby establishing the result.

In a more complete treatment, we must define suitable topologies for $C_K$
and $C_K^{SPIN}$ and derive equations like (7.10) as consequences of these
topologies. This task is carried out in the second paper of ref. 8.

There are several physical systems of interest other than point particles in
${\bf R}^d$ to which the techniques outlined here can be extended. It has
been shown elsewhere for example [2] that there exist exotic possibilities
for the statistics of strings in ${\bf R}^3$ if antistrings are ignored.
[These strings can be vortex rings in $He^4$ or strings produced in GUT's
during phase transitions.] A spin-statistics theorem for these strings as
well has been proved in ref. 9 by including antistrings and
creation-annihilation processes, and it will rule out these exotic
statistics.

\medskip\

\noindent {\bf Acknowledgement\addcontentsline{toc}{section}{Acknowledgement}%
}

This work was supported by the U.S. Department of Energy under Contract
Number DE-FG02-85ER40231.

\end{document}